\documentclass[12pt]{extarticle}
\usepackage[utf8]{inputenc}
\usepackage[margin=1in]{geometry}
\usepackage{mathtools}
\usepackage{booktabs}
\usepackage{amsmath}
\usepackage{appendix}
\usepackage{graphicx}
\usepackage[para]{threeparttable}
\usepackage[super]{natbib}
\usepackage{setspace}
\usepackage{lineno}
\usepackage{lipsum}
\usepackage{url}
\usepackage{hyperref}
\usepackage{adjustbox}
\usepackage{listings}
\usepackage{float}

\begin{document}

\title{\Large{\textbf{An Approximate Quasi-Likelihood Approach for Error-Prone Failure Time Outcomes and Exposures}}}
\author{Lillian A. Boe$^{1,*}$, 
Lesley F. Tinker$^{2}$, and 
Pamela A. Shaw$^{1}$ \\
$^{1}$Department of Biostatistics, Epidemiology, and Informatics, \\ University of Pennsylvania Perelman School of Medicine, \\  Philadelphia, PA 19104\\
$^{2}$WHI Clinical Coordinating Center,\\ Fred Hutchinson Cancer Research Center, Seattle, WA 98109 \\
\textit{*email}: boel@pennmedicine.upenn.edu}

\label{firstpage}

\date{}
\maketitle
\vspace{-0.5in}

\begin{center}

\end{center}

\begin{abstract}
Measurement error arises commonly in clinical research settings that rely on data from electronic health records or large observational cohorts. In particular, self-reported outcomes are typical in cohort studies for chronic diseases such as diabetes in order to avoid the burden of expensive diagnostic tests. Dietary intake, which is also commonly collected by self-report and subject to measurement error, is a major factor linked to diabetes and other chronic diseases.  These errors can bias exposure-disease associations that ultimately can mislead clinical decision-making. We have extended an existing semiparametric likelihood-based method for handling error-prone, discrete failure time outcomes to also address covariate error. We conduct an extensive numerical study to compare the proposed method to the naive approach that ignores measurement error in terms of bias and efficiency in the estimation of the regression parameter of interest. In all settings considered, the proposed method showed minimal bias and maintained coverage probability, thus outperforming the naive analysis which showed extreme bias and low coverage. This method is applied to data from the Women's Health Initiative to assess the association between energy and protein intake and the risk of incident diabetes mellitus. Our results show that correcting for errors in both the self-reported outcome and dietary exposures leads to considerably different hazard ratio estimates than those from analyses that ignore measurement error, which demonstrates the importance of correcting for both outcome and covariate error. Computational details and R code for implementing the proposed method are presented in Section \ref{Rcode} of the Supplementary Materials. 
\end{abstract}


\section{Introduction}

Chronic diseases are often recorded primarily by self-reported diagnosis in large observational cohort studies. For example, in comparison to reference (gold) standard measures for detecting diabetes, such as fasting glucose and hemoglobin A1c (HbA1c), self-reported diabetes status is inexpensive and easily attainable. However, not all people who are diagnosed with diabetes or other conditions will self-report that they have the disease. Reasons for failing to report having a chronic condition include failure to be diagnosed, lack of understanding about the disease, and a belief that the disease has gone away if it is being properly treated.\citep{centers2017national,shah2008self} Conversely, a positive disease status is occasionally reported when the disease is not actually present.\citep{ning2016comparison,schneider2012validity} Dietary intake, which is also commonly recorded by self-report, is thought to play a crucial role in determining the risk of chronic diseases such as diabetes and cardiovascular disease. In nutritional epidemiology, estimates of diet-disease associations can be distorted due to measurement error in both self-reported dietary exposures and disease outcomes. A new analytic approach is needed to properly relate error-prone exposures with error-prone disease outcomes of interest. In this paper, we have extended an existing semiparametric model for handling failure time outcomes assessed through interval-censored, error-prone measures to also address measurement error in the exposure variable.

There is ample literature available on methods for adjusting analyses with error-prone exposures in the case of time-to-event outcomes.\citep{carroll2006measurement} In existing epidemiological analyses, regression calibration is one of the most popular methods for addressing covariate measurement error.\citep{shaw2018citation}  This method relies on building a calibration model that relates the expected value of the unobserved true exposure to the observed data. \citet{prentice1982covariate} introduced the method for time-to-event outcomes.  Rosner et al. considered it for logistic regression, where a single or multiple covariates were error-prone.\cite{rosner1989correction,rosner1990correction} In non-linear models, such as Cox and logistic regression, regression calibration is considered a quasi-likelihood approach as it is generally only an approximate correction,\citep{buono10} but it has been observed to do well for modest $\beta$ and low event rates. \citep{prentice1982covariate, carroll2006measurement} The popularity of this approach likely has to do with the intuitive appeal of the method and the ease of implementation. The method proposed in this manuscript uses regression calibration in order to develop an estimator that will correct for both covariate and outcome error.

Compared to methods for addressing covariate error, there has been notably less investigation into methods that correct for errors that occur in the time-to-event outcomes themselves. In epidemiologic cohort studies, the time-to-event of interest is often ascertained through periodic follow-up, thus resulting in data captured in fixed intervals. Thus, methods that address errors in the event indicator at each interval are of particular interest. \citet{balasubramanian2003estimation} developed estimation methods for the distribution of the time-to-event that consider various periods of exposure and diagnostic tests with different levels of accuracy. \citet{meier2003discrete} presented an adjusted proportional hazards model for estimating hazard ratios in discrete time survival analysis when the outcome is measured with error. \citet{magaret2008incorporating} considered methods that adjusted the proportional hazards model to incorporate data from validation subsets for the case where the sensitivity and the specificity of the diagnostic tests are unknown. All of this existing work assumes that the covariates included in the time-to-event analyses are error-free, which is often untrue with clinical data.

This manuscript specifically builds on the work of Gu et al., \cite{gu2015semiparametric} which introduced a semiparametric likelihood-based approach for estimating the association of covariates with an error-prone discrete failure time outcome. Motivated by an example from the Women's Health Initiative (WHI), we extend this method to incorporate a regression calibration fix that additionally adjusts for covariate measurement error and also allows for strata-specific baseline hazards. Our method can be applied to a study cohort that has collected follow-up data on an error-prone disease status variable at two or more distinct visit times and has information available at baseline on specific covariates of interest. In the presence of covariate measurement error, the proposed method can be considered when there is data that informs the measurement error model. We must assume that (1) information is available regarding the sensitivity and specificity of the outcome measure (2) a second measure of the error-prone covariate(s) is available on at least a subset.

Section \ref{allmethods} introduces the theoretical development of the method by providing notation, constructing the likelihood function and discussing the proposed adjustment method that corrects for outcome and covariate error. Next, we examine the numerical performance of the proposed method with a simulation study in Section \ref{section3}. In Section \ref{section4}, we apply the proposed method to evaluate the association between dietary energy, protein, and protein density intake and incident diabetes in a subset of women enrolled in the WHI. Finally, we highlight the important findings of this work and discuss potential extensions in Section \ref{discussion}. 
\section{Methods}\label{allmethods}
\subsection{Notation and Time-to-Event Model}\label{ref_section1} 
Let $T_i$ be the unobserved time-to-event of interest for subjects $i=1,...,N$. Consider a study with periodic follow-up where each subject may have a slightly different visit schedule or missed visits. Define $\tau_1,...,\tau_J$ as the distinct possible visit times among all $N$ subjects. Denote $\tau_0=0$ and $\tau_{J+1}=\infty$. We assume that the time to the event of interest is continuous, but follow-up occurs at discrete visit times. The follow-up time period can then be divided into $J+1$ disjoint intervals, listed as follows: $[\tau_0,\tau_1),[\tau_1,\tau_2),...[\tau_J,\tau_{J+1})$. Assume that all subjects in the study are event-free at time $\tau_0$. Later, we will relax this assumption. Let $n_i$ be the number of visits for the $i^{th}$ subject, which we assume is random. In our motivating data example, each subject self-reports his or her disease status at each visit time, potentially with error, up until the first positive. Our method can also be applied to the more general setting for error-prone outcomes in which follow-up continues beyond the first positive. Define $\mathbf{Y_i}$ and $\mathbf{t_i}$ as the random $1\times n_i$ vector of error-prone outcomes and corresponding vector of visit times for subject $i$. Specifically, define $Y_{ij}$ as 1 if the $j^{th}$ error-prone outcome for $i^{th}$ subject is positive, and 0 otherwise. Then, the joint probability of the observed data for the $i^{th}$ subject is:
\begin{equation}\label{firstequation}
    f(\mathbf{Y_i},\mathbf{t_i},n_i)=\sum_{j=1}^{J+1}\theta_j\Pr(\mathbf{Y_i},\mathbf{t_i},n_i|\tau_{j-1}< T_i\leq \tau_j),
\end{equation}
\noindent where $\theta_j=\Pr(\tau_{j-1}<T_i\leq\tau_j)$.

We make the additional assumption that conditioned on the true event time $T_i$, the $n_i$ error-prone outcomes $Y_{ij}$ are independent, i.e. $\Pr(\mathbf{Y_i}|T_i,\mathbf{t_i})=\prod_{l=1}^{n_i}\Pr(Y_{il}|T_i,t_{il})$. Thus, other observed error-prone outcomes do not provide additional information about a specific error-prone outcome beyond what is already given by the true time of event. Following the notation and logic of Balasubramanian and Lagakos, \cite{balasubramanian2003estimation} it can be shown that for the case of a prespecified visit schedule, the likelihood becomes:
\begin{equation}\label{jointprob}
    f(\mathbf{Y_i},\mathbf{t_i},n_i)=\sum_{j=1}^{J+1}\theta_j\left[\prod_{l=1}^{n_i}\Pr(Y_{il}|\tau_{j-1}<T_i\leq \tau_j,t_l)\right]=\sum_{j=1}^{J+1}\theta_jC_{ij},
\end{equation}
\noindent where $C_{ij}=\prod_{l=1}^{n_i}\Pr(Y_{il}|\tau_{j-1}<T_i\leq \tau_j,t_l).$ In Section S2 of the Supplementary Materials, we show how equation (\ref{jointprob}) becomes the expression for a subject's likelihood contribution.

For ease of presentation, we calculate $C_{ij}$ for the case of no missed visits, but the formula can be easily adapted to accommodate missed visits by summing up the $\theta_j$ for the $(\tau_{j-1},\tau_j]$ that define each subject's observational interval. We assume constant and known sensitivity $(Se)$ and specificity $(Sp)$; namely, $Se=\Pr(Y_{il}=1|\tau_{j-1}<T_i\leq\tau_j,t_l\geq\tau_j)$ and $Sp=\Pr(Y_{il}=0|\tau_{j-1}<T_i\leq\tau_j,t_l\leq\tau_{j-1})$.
\noindent Then, the $C_{ij}$ terms take the following form:
\begin{align*}
    \begin{matrix}  C_{i1}=Se^{\sum_{j=1}^{n_i}Y_{ij}}(1-Se)^{\sum_{j=1}^{n_i}(1-Y_{ij})},\\  C_{i2}=Sp^{(1-Y_{i1})}(1-Sp)^{Y_{i1}}Se^{\sum_{j=2}^{n_i}Y_{ij}}(1-Se)^{\sum_{j=2}^{n_i}(1-Y_{ij})}, \\ 
 ... \\
    C_{i(J+1)}=Sp^{\sum_{j=1}^{n_i}(1-Y_{ij})}(1-Sp)^{\sum_{j=1}^{n_i}Y_{ij}}. \end{matrix}
\end{align*}
Now suppose we have the proportional hazards model, $S(t)=S_0(t)^{\exp(x^T\beta_X+z^T\beta_Z)}$. We assume that one or more covariates are recorded with error. Define $X_i^*$ as a $p-$dimensional vector of covariates of interest that may be observed with error, and $X_i$ a corresponding $p-$dimensional vector of unobserved true exposure variables. We describe the error structure of the observed error-prone covariate $X^{*}$ in Section \ref{proposed}. Let $Z_i$ be a $q-$dimensional vector of precisely measured covariates  (i.e. error-free) that may be correlated with $X_i$. Define $\beta=(\beta_X,\beta_Z)^T$. The likelihood can be rewritten in terms of the baseline survival probabilities $\mathbf{S}=(S_1,S_2,...,S_{J+1})^T$ defined by the random variable $T_0$ with survival function $S_0(t)$, where $S_j=\Pr(T_0>\tau_{j-1})$. One then has $1=S_1>S_2>...>S_{J+1}>0$ and $S_j=\sum_{h=j}^{J+1}\theta_{h}$. We can define a linear $(J+1)\times(J+1)$ transformation matrix $M$ such that $\mathbf{\theta}=M\mathbf{S}$. Finally, define the $(N)\times(J+1)$ matrix $D=CM$, where $C_{N\times(J+1)}$ consists of the $C_{ij}$ elements defined above. Following Gu et al., \cite{gu2015semiparametric} the log-likelihood can be rewritten as:

\begin{equation}\label{likelihoodNoCov}
     l(\mathbf{S},\beta)=\sum_{i=1}^{N}\log\left(\sum_{j=1}^{J+1}D_{ij}S^{(i)}_j\right),
\end{equation}

\noindent where $S_j^{(i)}=(S_j)^{\exp(x_i^T\beta_X+z_i^T\beta_Z)}$. Thus, the log-likelihood can now be rewritten as:
\begin{equation}\label{likelihood}
     l(\mathbf{S},\beta)=\sum_{i=1}^{N}\log\left(\sum_{j=1}^{J+1}D_{ij}S_j^{\exp(x_i^T\beta_X+z_i^T\beta_Z)}\right).
\end{equation}

The $D_{ij}$ components of the log-likelihood consist of elements of the matrix $D$ defined above and are functions of the observed data, $(X_i, Z_i, Y_i, t_i)$, as well as $Se$ and $Sp$. One can apply the usual  maximum likelihood approach to solve for the unknown parameters $\beta_X$, $\beta_Z$, $S_2$,...,$S_{J+1}$. The covariance matrix can be found by inverting the Hessian matrix. Note that the model above introduced by \citet{gu2015semiparametric} is considered semiparametric because we do not make any assumptions about the form of the baseline survival probabilities, $S_j$, for $j=1,...,J+1$.

\subsection{Proposed Method for Outcome and Covariate Error}\label{ref_section2}

We now extend the above method that corrects for outcome error in the discrete proportional hazards model  to also adjust for covariate error by adopting a regression calibration type approach. In this section, we describe the regression calibration approach for covariate measurement error, present our proposed method to adjust for covariate and outcome error, extend our method to accommodate a baseline hazard that varies across strata, and extend the method to handle false negatives that are mistakenly included in the analysis.

\subsubsection{Regression calibration for covariate error}
Regression calibration is an approach to correcting biases in regression parameters when exposure variables are recorded with error, in which a calibration equation for the unobserved exposure $X$ is estimated.  Namely, one builds a model for $E(X|X^*,Z)$, where $X^*$ is the error-prone observation or surrogate for $X$ while $Z$ are the other precisely observed covariates in the outcome model (\ref{likelihood}). Regression calibration may be used when $X^*$ follows the classical measurement error model or when $X^*$ has both systematic and random error. These error settings will be explained in further detail in the subsequent section.  \citet{rosner1989correction} introduced a post-hoc calibration fix in the logistic regression setting when there is measurement error in a single covariate of interest and \citet{rosner1990correction}  extended the method to handle  multiple error-prone covariates in logistic regression. In each of these approaches, the calibration equation is used to correct the naive parameter estimates that are obtained from first fitting the outcome regression that ignores the measurement error. An asymptotic formula for the variance that incorporates the uncertainty of the calibration equation is derived using the Delta method. We will employ a similar post-hoc calibration fix-up for the estimator that first corrects for outcome measurement error. We further justify why this post-hoc correction approach is expected to work well in our discrete-time proportional hazards setting at the end of this section. 

\subsubsection{Proposed approach for outcome and covariate error}\label{proposed}

Recall that $X_i$ is a $p-$dimensional vector of true, unobserved covariates, while $Z_i$ is a $q-$dimensional vector of observed, precisely measured covariates possibly correlated with $X_i$. Instead of observing $X_i$, we assume an error-prone $X_i^*$ is observed, where $X_i^*$ is assumed to be linearly related with $X_i$ and possibly other covariates $Z_i$. This error model has been commonly applied in many settings, including nutritional epidemiology.\citep{carroll2006measurement,keogh20} The regression calibration model then takes the following form:
\begin{equation}\label{regcalmodel}
    X_i=\delta_{(0)}+\delta_{(1)}X^{*}_i+\delta_{(2)}Z_i+U_i,
\end{equation}

\noindent where $U_i$ is a random, mean 0 error term, which is independent of $X^*_i$ and $Z_i$.  Equation (\ref{regcalmodel}) directly implies that our observed, error-prone variable $X_i^{*}$ follows the linear measurement error model, i.e. $X_i^*=\alpha_{(0)}+\alpha_{(1)}X_i+\alpha_{(2)}Z_i+e_i$, where the random error $e_i$ is independent of $X_i$ and $Z_i$.\citep{keogh20} Note that we also assume non-differential error, i.e. the distribution of $T$ conditional on $(X,X^*,Z)$ is equal to the distribution of $T$ conditional on $(X,Z)$. The model parameters in equation (\ref{regcalmodel}) are identifiable if we have a calibration subset available in which we observe the error-prone measure $X_i^{*}$, as well as a measure $X_i^{**}$ that is  unbiased for the true $X_i$ and follows the classical measurement error model: 
\begin{equation}\label{classicalme}
X_i^{**}=X_i+\epsilon_i,
\end{equation}

\noindent where $\epsilon_i$ is random, mean 0 error that is independent of $X_i$. $X^{**}$ is often referred to as an imperfect reference or alloyed gold standard.\cite{shaw20,spiegelman97} Note that $\epsilon_i$ are assumed to be independent of all variables in the outcome model (\ref{likelihood}). Observing the exact true exposure $X_i$ in the ancillary data is a special case of observing $X_i^{**}$ where the error variance is 0, and the subset is typically called a validation subset. A special case of the linear measurement error model occurs when $\alpha_{(0)}=\alpha_{(2)}=0$ and $\alpha_{(1)}=1$, and thus the observed error-prone measurement $X^*_i$ has classical measurement error. In this scenario, we can estimate the parameters of the calibration model by assuming that we observe replicates of $X^*$. Ancillary data of this type is typically referred to as a reliability subset. 

When a calibration or validation subset is available, one can adopt a regression calibration type approach to further correct the regression coefficients for error in the exposure variable. In the case of a calibration subset, we regress $X_i^{**}$ on the error-prone exposure, $X_i^*$, and other covariates of interest $Z_i$ to fit the model:
\begin{equation}\label{xdoublestar}
    X^{**}_i=\delta_{(0)}+\delta_{(1)}X_i^{*}+\delta_{(2)}Z_i+V_i,
\end{equation}

\noindent where $V_i$, is random, mean 0 error. Note the model in equation (\ref{xdoublestar}) differs from that in equation (\ref{regcalmodel}) only in that the error term $V_i$ incorporates the extra variability introduced by the error term in $X_i^{**}$. Estimates of the coefficients from fitting this linear regression can then be used to correct the $\beta$ coefficients from the time-to-event model. Following the approach of Rosner et al., \cite{rosner1990correction} the corrected $\beta$ can be found by solving:
\begin{equation}\label{posthoccorr}
    \hat{\beta}=\hat{\beta}^*\hat{\Delta}^{-1},
\end{equation}
\noindent where $\hat{\beta}^*$ is the partially ``naive" regression coefficient obtained from the time-to-event model ignoring the error in $X^*$, and $\hat{\Delta}$, the estimated multivariate correction factor, is defined as:
\begin{equation}\label{multicorrection}
\hat{\Delta}=\begin{bmatrix}
\hat{\delta}_{(1)p\times p} & \hat{\delta}_{(2)p\times q} \\
0_{q\times p} & I_{q\times q} 
\end{bmatrix}_.
\end{equation}
The variance-covariance matrix $\Sigma$ for $\hat{\beta}$ is calculated using the multivariate delta method. Assume that $\hat{\beta}^*$ and $\hat{\Delta}$ are independent, which  holds if the calibration subset is an independent group of individuals from the main study (i.e. the main study data and the calibration subset are either independent data sets or are mutually exclusive subsets of the same set of data) and approximately holds if  the number of subjects in the calibration subset, $n_c$, is a small percentage of the main study sample size, $N$.\citep{rosner1989correction} Once we make this connection, we see that we can apply the same formulas as \citet{rosner1990correction} and therefore the $(j_1,j_2)^{th}$ element of $\hat{\Sigma}$ for $\hat{\beta}$ is
\begin{equation}
   \hat{\Sigma}_{\beta}(j_1,j_2)\cong\left(\hat{A}'\hat{\Sigma}_{\beta^*}\hat{A}\right)_{j_1,j_2}+\hat{\beta}^*\hat{\Sigma}_{A,j_1,j_2}\hat{\beta}^{*'}, 
\end{equation}
\noindent where $\hat{A}=\hat{\Delta}^{-1}$, $\hat{\Sigma}_{\beta^*}$ is the corresponding estimated variance-covariance matrix, and $\hat{\Sigma}_{A,j_1,j_2}$ is described below. Note that $\hat{\Sigma}_{\beta^*}$ can be estimated from the model introduced above that only adjusts for outcome error. The matrix $\hat{\Sigma}_{\beta}(j_1,j_2)$ is essentially a sum of two pieces: the first can be viewed as the contribution of the uncertainty in estimating $\beta^*$ and the second is a contribution of the uncertainty in the calibration coefficients.
Following Rosner et al., \cite{rosner1990correction} the $(i_1,i_2)^{th}$ element of $\hat{\Sigma}_{A,j_1,j_2}$, for $i_1,i_2,j_1,j_2=1,...,w, (w=p+q)$ is
\begin{equation}
    \hat{\Sigma}_{A,j_1,j_2}\cong\sum_{r=1}^w\sum_{s=1}^w\sum_{t=1}^w\sum_{u=1}^w\hat{A}_{i_1r}\hat{A}_{sj_1}\hat{A}_{i_2t}\hat{A}_{uj_2}Cov(\hat{\Delta}_{rs},\hat{\Delta}_{tu}).
\end{equation}

In the simple linear regression case, the post-hoc correction presented in equation (\ref{posthoccorr}) reduces to the following familiar form: $
    \hat{\beta}=\frac{\hat{\beta}^*}{\hat{\delta}}$, where $\hat{\beta}^*$ is the estimate for $\beta$ obtained from the ``naive" regression using $X_i^*$ that ignores the error in the covariate of interest, and $\hat{\delta}$ is the estimate of the attenuation coefficient from the simple linear regression correction. Similarly, the variance estimator for this correction is easily calculated using the univariate delta method as
    $var(\hat{\beta})=\frac{1}{\hat{\delta}^2}var(\hat{\beta}^*)+\frac{\hat{\beta}^{*^2}}{\hat{\delta}^4}var(\hat{\delta})$.
    
 \citet{rosner1990correction} justified this proposed correction for logistic regression for small $\beta$. One can use a Taylor series approximation to show when this method can be expected to work similarly for the Cox proportional hazards model. Specifically, \citet{green1983comparison} used a linear Taylor series expansion to illustrate the approximate mathematical equivalence between the logistic regression model and the Cox proportional hazards model when the event of interest is rare, the follow-up time is short, and the baseline hazard in the Cox model is constant. The post-hoc regression parameter correction developed for logistic regression is expected to do similarly well for the Cox proportional hazards model for settings that uphold these assumptions. We explore this further with a numerical study. In Section S3 of the Supplementary Materials, we establish the asymptotic properties of our estimator.

\subsubsection{Strata-specific baseline hazards}

For a continuous failure time outcome, the proportional hazards model takes the familiar form $S(t)=S_0(t)^{\exp(x^T\beta_X+z^T\beta_Z)}$. Under this assumption, the baseline survival function $S_0(t)$ and baseline hazard function $\lambda_0(t)$ are shared by all subjects in the data. Oftentimes, however, this assumption is invalid and we expect baseline survival to differ across groups defined by one or more covariates. To address the issue of non-proportional hazards, we let the survival function for a subject from stratum $k$ be $S_k(t)=S_{0k}(t)^{\exp(x^T\beta_X+z^T\beta_Z)}$, $k=1,...,K$, where $S_{0k}(t)$ is the baseline survival for all individuals in stratum $k$.

In a discrete proportional hazards model that incorporates stratification, we allow strata-specific versions of the baseline survival function introduced in Section \ref{ref_section1}, such that $\mathbf{S}_k=(S_{1k},S_{2k},...,S_{(J+1)k})^T$. We can accordingly modify the log-likelihood function from equation (\ref{likelihood}) to allow for stratification on one or more predictors. As in the continuous time setting, the stratified log-likelihood for all $N$ subjects is a simple sum of the log likelihood for each stratum. Now, in our discrete failure time setting, the log-likelihood function for the $N_k$ subjects in stratum $k$ is given by:
\begin{equation}
    l_k(\mathbf{S}_k,\beta)=\sum_{i=1}^{N_k}\log\left(\sum_{j=1}^{J+1}D_{ij}S_{jk}^{\exp(x_i^T\beta_X+z_i^T\beta_Z)}\right).
\end{equation}
Correspondingly, the log likelihood for all $N$ subjects is calculated as follows: 
\begin{equation}\label{stratlik}
    l(\mathbf{S}_k,\beta)=\sum_{k=1}^{K}\left[\sum_{i=1}^{N_k}\log\left(\sum_{j=1}^{J+1}D_{ij}S_{jk}^{\exp(x_i^T\beta_X+z_i^T\beta_Z)}\right)\right].
\end{equation}
Using this likelihood, we can solve for the unknown parameters $\beta_X$, $\beta_Z$, $S_{2k}$,...,$S_{(J+1)k}$, $k=1,...,K$ and compute the estimated covariance matrix as described in Section \ref{ref_section1}. Although the baseline survival functions are different for each stratum, the coefficients $\beta_X$ and $\beta_Z$ are assumed to be uniform across all strata. Note, in the setting without misclassification in the event indicator, strata should be chosen such that each stratum contains subjects with the event of interest, as a stratum with no events does not contribute any information to the analysis.\citep{harrell2015regression} However, with a sensitivity less than 1, events and non-events of a stratum both contribute to the likelihood (\ref{stratlik}). Under this model, we can apply the same post-hoc fix introduced in Section \ref{proposed} to also correct the estimated coefficients for exposure error. 

\subsubsection{Adjusting for false negatives at baseline}\label{falsenegs}

The proposed method can be modified to handle the case in which individuals with a baseline false negative test are erroneously included into the analysis. This simple extension of the method applies to scenarios in which subjects are only included in the study if they report being event-free at baseline. This extension is motivated by the analysis approach of Tinker et al., \cite{tinker2011biomarker} which excluded anyone with a positive self-report at baseline. To allow for a non-zero probability of a baseline false negative test, we will now assume $S_1<1$. 

Let $R_i$ and $E_i$ be the observed error-prone event status at baseline and the unobserved true event status at baseline, respectively. Consider all subjects in the study that have a negative error-prone outcome at baseline, i.e. $R_i=0$, and are therefore included in the analysis population. Define $\eta$ as the negative predictive value, or the probability that a subject with a negative error-prone outcome is truly disease-free, i.e. $\eta=\Pr(E_i=0|R_i=0)$, which we assume is constant across all $N$ subjects. Further assume all subjects with a negative error-prone outcome who are truly disease-free constitute a random sample of all subjects who are truly disease-free at baseline, so that $\Pr(\mathbf{Y_i},\mathbf{t_i},n_i|E_i=0,R_i=0)=\Pr(\mathbf{Y_i},\mathbf{t_i},n_i|E_i=0)$. Then, the likelihood function for subject $i$ can be expressed as follows:
\begin{eqnarray*}
    f(\mathbf{Y_i},\mathbf{t_i},n_i)&=&\Pr(\mathbf{Y_i},\mathbf{t_i},n_i|R_i=0) \\
    &=&\eta\Pr(\mathbf{Y_i},\mathbf{t_i},n_i|E_i=0,R_i=0)+(1-\eta)\Pr(\mathbf{Y_i},\mathbf{t_i},n_i|E_i=1,R_i=0)\\
  & =&\eta\sum_{j=1}^{J+1}D_{ij}S_j^{\exp(x_i^T\beta_X+z_i^T\beta_Z)}+(1-\eta)D_{i1}S_1^{\exp(x_i^T\beta_X+z_i^T\beta_Z)}.
\end{eqnarray*}

\noindent Thus, the log-likelihood for all $N$ subjects is 
\begin{equation}
    l(\mathbf{S},\beta)=\sum_{i=1}^{N}\log\left(D_{i1}S_1^{\exp(x_i^T\beta_X+z_i^T\beta_Z)}+\eta\sum_{j>1}^{J+1}D_{ij}S_j^{\exp(x_i^T\beta_X+z_i^T\beta_Z)}\right).
\end{equation}

\section{Numerical Study}\label{section3}

We examine the numerical performance of our proposed estimator using a simulation study. We compare our estimator to the results from the ``true" model, in which a discrete proportional hazards model is fit with the true (error-free) event time and covariate values, and the ``naive" model, which fits the same model with the error-prone outcome and covariate.  In all simulations, we assume a single error-prone covariate of interest. We assume that there are two precisely measured covariates, which are moderately correlated with the error-prone variable. Our results show how our estimator performs under different levels of outcome sensitivity and specificity, error variance in the covariate, sample size, and censoring rates. We present percent biases, average standard errors (ASE), empirical standard errors (ESE), and 95\% coverage probabilities (CP) across these various settings. Mean percent bias is calculated as follows: $\frac{\hat{\beta}-\beta}{\beta}\times100$, where $\beta$ is the target regression parameter of interest. The ASE is defined as the mean of the estimated standard errors from the model, while the ESE is the empirical standard deviation of the estimated coefficients across simulations. Additionally, we present type I error results for $\beta_{X1}=0$ and $\alpha=0.05$, where $\beta_{X1}$ is the regression parameter corresponding to the error-prone covariate.

\subsection{Simulation Setup}

We present results from 1000 simulations run in R version 3.5.2.\citep{citeR} The three covariates, $X_1$, $Z_1$, and $Z_2$ were generated from a multivariate normal distribution, all with mean 0 and a covariance matrix with all diagonal elements equal to 1 and all off-diagonal elements equal to 0.3. We generated our error-prone covariate $X_1^*$ using the linear measurement error model, $X_1^*=\alpha_0+\alpha_1X_1+\alpha_2Z_1+\alpha_3Z_2+e$, with $\alpha_0=1$, $\alpha_1=0.8$, $\alpha_2=0.3$, and $\alpha_3=0.5$. We assumed $e \sim N(0,\sigma^2)$  and considered $\sigma^2$ values of 0.59 and 1.72, which correspond to estimated $\delta_{(1)}$ values of approximately $0.60$ and $0.30$, respectively. 

Later, we assess how our method performs when error is not normally distributed, but instead $e\sim .4\mathcal{N}(0, 1) + .6\mathcal{N}(2, 1.5)$ and $e$ distributed as a \textit{t} with 4 degrees of freedom (df). For all simulations, there are $N=1000$ subjects in the main study data. We assume our calibration subset is a random sample of $n_C=500$ subjects from the main study. The measure approximating $X_1$ in the calibration subset, $X^{**}_1$, is generated to follow the classical measurement error model from equation (\ref{classicalme}), where $\epsilon \sim \mathcal{N}(0,0.06)$.

%\begin{center}
%$\Sigma=\begin{bmatrix}
%1 & 0.3 & 0.3 \\
%0.3 & 1 & 0.3 \\
%0.3 & 0.3 & 1
%\end{bmatrix}_.$
%\end{center}

We considered typical settings for which regression calibration has been observed to perform well, including a moderate $\beta_{X1}$ and a higher censoring rate.\citep{shaw12}
The true log hazard ratios were selected to be $\beta_{X1}=\log(1.5)$, $\beta_{Z1}=\log(0.7)$, and $\beta_{Z2}=\log(1.3)$. Later, we set $\beta_{X1}=\log(3)$ to assess how the method performs under a more extreme regression coefficient corresponding to the error-prone covariate. The true time-to-event was generated from a continuous time exponential distribution. To mimic the settings of real data, we considered a follow-up schedule with four possible visit times. To obtain an average true censoring rate (CR) of approximately 0.90, we set the visit times to be $\{2,5,7,8\}$ with baseline hazard rates of 0.012 and 0.008 for $\beta_{X1}=\log(1.5)$ and $\beta_{X1}=\log(3)$, respectively. Fixing the visit times at $\{1,3,4,6\}$  and baseline hazard rates at  0.094 and 0.076 for $\beta_{X1}=\log(1.5)$ and $\beta_{X1}=\log(3)$, respectively, leads to an average true CR of approximately 0.55. Note that the visit times are not required to be equally spaced. Figure \ref{fig:my_label} in the Supplementary Materials depicts the estimated nonparametric maximum likelihood estimators of the survival distribution for the true and error-prone outcomes under the two CRs for $\beta_{X1}=\log(1.5)$  for a single simulated data set. 

To assess how our method performs when the baseline hazard varies across strata, we simulate four approximately equal sized strata. For test times at $\{2,5,7,8\}$, we let the four baseline hazard rates be 0.008, 0.010, 0.011, and 0.019, which resulted in an overall censoring rate of approximately $90\%$. Similarly, to obtain an overall censoring rate of approximately $55\%$, the baseline hazard rates for each stratum were fixed at 0.090, 0.080, 0.075, and 0.131 for visit times at $\{1,3,4,6\}$.

 To capture the interval in which each simulated event occurred, we created an indicator for whether or not the current visit time was greater than the actual event time itself. This indicator variable was ``corrupted" using sensitivity and specificity values in order to create the error-prone vector of outcomes, $\mathbf{Y_i}$. To mimic a diagnostic test with different levels of accuracy, we considered the case where sensitivity $=0.90$ while specificity $=0.80$, and sensitivity $=0.80$ while specificity $=0.90$. Later, we assess the performance of the proposed method when a baseline negative predictive value ($\eta$) less than 1 is incorporated into the analyses to adjust for erroneously included false negative participants. We vary $\eta$ between 0.98 and 0.90. To simulate this scenario, we set the true time-to-event equal to 0 for a fixed proportion of subjects, $\eta$, included in the data. This represents an event time prior to the start of the study. Additionally, we show that the proposed method can handle different visit structures by allowing each visit to be subject to a constant, independent probability of missingness, which mimics the Missing Completely at Random (MCAR) setting for missing data. To simulate this, we create a binary variable indicating whether the $j$th visit is missing for each subject using a fixed probability $P_{Miss}$ of either 0.10 or 0.40. We further assess the method under parameters that mimic the structure of the WHI data example, with $N=65,000$, $n_C=500$, $Se=0.61$ and $Sp=0.995$, $\eta=0.96$, and a censoring rate of 95\% for the error-prone discrete failure time. To simulate self-reported outcomes in the WHI data, we stopped visit times for each subject after the first positive error-prone outcome. Regression coefficients for the discrete time Cox proportional hazards model and the true data can be estimated by fitting a generalized linear model assuming the binomial outcome and complementary log-log link.\citep{hashimoto2011regression}

\subsection{Simulation Results}
Tables \ref{table1sims}-\ref{table4sims} present estimates of mean percent bias, ASE, ESE, and 95\% CP across the various settings described above. For Table \ref{table1sims}, we consider the case where $\beta_{X1}=\log(1.5)$. Overall, we see that the proposed method improves over naive analyses in bias and in the nominal coverage of 95\% confidence intervals. In fact, under various different settings, the percent bias of our parameters of interest never exceeds 5\%. Additionally, we maintain nominal coverage for a 95\% confidence interval. Furthermore, our ASEs closely resemble the ESEs, demonstrating that our standard error estimates also performed well. In contrast, for the analyses that ignore measurement error, estimates of $\beta_{X1}$, $\beta_{Z1}$, and $\beta_{Z2}$ have bias as high as $-96.33\%$ and attain very little coverage. Table \ref{table1simsappendix} in the Supplementary Materials further shows results for the method that corrects for covariate error only and the method that corrects for outcome error only under these same simulation settings. Regression parameters for the method correcting for only covariate error have absolute mean percent biases ranging from $47.05$ to $82.31$, while the method correcting solely outcome error has bias ranging from $5.840$ to $70.28$. Unsurprisingly, the proposed method greatly improves over all three alternative approaches that ignore measurement error to some degree.

In Table \ref{table2sims}, we set $\beta_{X1}=\log(3)$. The method still performs reasonably well when the censoring rate is high (CR = $0.90$), as absolute percent bias stays below 12\% and nominal coverage is maintained. However, when the censoring rate decreases to $0.55$, we begin to see an increase in bias and a steep decrease in coverage, particularly for $\beta_{X1}$. This is unsurprising, as regression calibration is known to break down with a larger $\beta$ coefficient and a higher event rate.\citep{shaw12} We observe that even in the most challenging scenarios for the proposed method, i.e. a more extreme $\beta_{X1}$, less censoring, and more covariate measurement error, the percent attenuation bias (coverage) was $17\%$ $(77\%)$ compared to $91\% (0\%)$ for the naive analysis.

In Table \ref{table3sims}, we examine the relative performance of our proposed method when the error in $X^*$ no longer follows a normal distribution. Here, we let the error in $X^*$ follow either a \textit{t} distribution with 4 df, or a mixture of two normals, as described in the simulation setup. On average, we observe $\delta_{(1)}=0.27$ when the error in $X^*$ follows the \textit{t} distribution and  $\delta_{(1)}=0.21$ when the error follows the mixture distribution, which reflects substantial error in our simulated covariate of interest in all scenarios. Since the applied regression calibration method assumes a first order approximation to estimate $E(X|X^*,Z)$, we expect the proposed method to perform best when the error in $X^*$ is normally distributed. Thus, it is unsurprising that the mean percent bias for the proposed method is a bit higher for $\beta_{X1}$ under these settings, particularly when the error follows a \textit{t} distribution. Nonetheless, absolute percent bias stays under 4\% in all scenarios. Most intervals still come very close to achieving the nominal level of 95\% CP. Our proposed approach still outperforms the naive method, which again shows severe bias of up to $-97.65\%$ and poor coverage. 

Table \ref{table4sims} shows the performance of the proposed method alongside the naive method in terms of mean percent bias, ASE, ESE, and 95\% CP when both approaches allow for stratification. In this table, we revert to letting the error in $X$ follow a normal distribution and set $\beta_{X1}=\log(1.5)$. We assume that there are four equally-sized strata. Similarly to what we observed in Table \ref{table1sims}, we see that the method performs well in terms of bias and coverage. Absolute bias for $\beta_{X1}$, corresponding to the error-prone covariate, ranges from $0.310\%$ to $1.893\%$ and is therefore quite low in all scenarios. The standard error estimator works well, as indicated by the attainment of nominal coverage. Again, we see extremely high bias for the naive approach, ranging from $-68.31\%$ to $-96.68\%$ for $\beta_{X1}$. 

Type I error results for the coefficient corresponding to the error-prone covariate are presented in Table \ref{T1errtab}. Type I error values ranged from 0.039 to 0.058 across different values of $Se$, $Sp$, $\delta_{(1)}$, and CR. With 1000 simulations, a 95\% confidence interval based on the true error rate $\alpha = 0.05$ is $(0.036,0.064)$. All calculated error rates in Table \ref{T1errtab} are within simulation error of the truth, indicating that type I error is preserved in the proposed method  for all settings.

Table \ref{supp2NPV} of the Supplementary Materials demonstrates the performance of the proposed method, now including adjustment for an imperfect baseline negative predictive value. Under different levels of covariate error and changes to the sensitivity and specificity, the bias of our parameters remains under 6\% and nominal coverage for a 95\% confidence interval is maintained, illustrating that the method performs well. We observe that the performance of the proposed method surpasses that of the naive method, which shows excessive bias in the parameters of interest, ranging from $-79.01\%$ to $-97.33\%$.

In Table \ref{supp3NPV} of the Supplementary Materials, we show that the proposed method can accommodate missed visits. Our approach performs well in all scenarios, maintaining an absolute mean percent bias of under $4.353\%$ when we let each visit to be subject to either 10\% or 40\%  missingness.  When there are missed visits, the proposed method outperforms the naive method, which shows extreme mean percent bias of up to $-96.85\%$.

Finally, we present results for the simulations that mimic the structure of the WHI data in Table \ref{WHIsims} of the Supplementary Materials. We see that the proposed method works well under measurement error settings similar to that of the WHI, maintaining an absolute percent bias of under $0.8\%$ for all scenarios. Again, the proposed method outperforms the naive method, in which we see absolute percent bias as high as $89.53\%$ for the regression parameter of interest and $0\%$ coverage probability for many scenarios. Similarly, the methods that correct for covariate error only and outcome error only both show extreme bias and inadequate coverage under these settings. 

\section{Women's Health Initiative (WHI) Example}\label{section4}

\subsection{WHI Study}

The Women's Health Initiative is a collection of studies launched in 1993 that together investigated the major causes of morbidity and mortality in US post-menopausal women.\citep{study1998design} We seek to examine the association between energy, protein and protein density (percentage of energy from protein) intakes with the risk of diabetes when all three exposures as well as diabetes status are self-reported and subject to error.\citep{neuhouser2008use,gu2015semiparametric} We analyze data on post-menopausal women aged 50-79 who participated in either the comparison arm of the Dietary Modification trial (DM-C) or the Observational Study (OS) and who had an average follow-up of approximately 9 years.\citep{ritenbaugh2003women,langer2003women} Neither women from the DM-C nor the OS received study interventions. The WHI also included the nutritional biomarker study which collected objective recovery biomarkers for energy and protein intake, thought to have only classical measurement error, on a subset of participants $(n_C=544)$. These biomarkers were previously used to develop calibration equations for the self-reported intakes of energy, protein and protein density.\citep{neuhouser2008use} Using these calibration equations, \citet{tinker2011biomarker} reported incident diabetes hazard ratios in this cohort for energy, protein, and protein density that were  corrected for the error in self-reported dietary exposures. Self-reported diabetes in the WHI has been reported to be subject to error.\citep{margolis2008validity} We apply our proposed method to correct for error in both the exposure and the diabetes failure time outcome. Our goal was to answer a similar research question as Tinker et al., \cite{tinker2011biomarker} only to use our method that additionally adjusts for error in the diabetes outcome. We adopted the same exclusion criteria as \citet{tinker2011biomarker} in order to arrive at our final analytic data set of 65,358 participants. In short, these criteria attempt to align the characteristics of DM-C and OS cohorts and exclude those with missing data or who reported diabetes at baseline. Baseline was defined as the time of the first self-reported dietary assessment post-enrollment, year 1 for the DM-C and year 3 for the OS. Further details are provided in Section S4 of the Supplementary Materials.  

We started with the previously developed calibration equations for dietary energy, protein, and protein density  from Neuhouser et al., \cite{neuhouser2008use} which we call our ``base" calibrations. Body mass index (BMI), age, race-ethnicity, income, and physical activity were included in the energy calibration model; BMI, age, race-ethnicity, income, and education for protein; and BMI, age, and smoking status for protein density. To avoid bias, regression calibration requires the calibration model to include the same covariates as the outcome model.\citep{rosner1990correction,kipnis2009modeling} We only considered the form of regression calibration in which the variables in the calibration and outcome models are exactly aligned. Thus, we extended each base calibration to include all predictors from our outcome model. Specifically, education, hypertension, and alcohol use were added to all calibrations. For each of the three nutrients, the calibration equation was fit by regressing the biomarker value $(X^{**})$ on the corresponding self-reported value and participant characteristics, as described above. 

In the WHI, prevalent diabetes was recorded via a self-reported questionnaire at baseline. We consider data from 8 years of annual follow-up visits in our analyses. Only the censored event-time was recorded in continuous time in our analytical dataset. Thus, we discretized the available data by dividing the follow-up time into 9 possible intervals. Then, for all 65,358 women in our analytic cohort, we considered the time at which the first occurrence of self-reported diabetes or censoring time was recorded and assumed that the occurrence of the censored self-reported outcome happened in the annual interval that the event time fell into. We note that in other settings our method could accommodate an increase in the number of time intervals if follow-up occurred more frequently than once a year (e.g. a bi-annual visit structure). 

Self-reported diabetes in the WHI was previously reported to have a sensitivity of 0.61, specificity of 0.995, and a baseline negative predictive value of 0.96.\citep{gu2015semiparametric}  We incorporated these values into our analyses. We also considered a sensitivity analysis in which we examined the results for a negative predictive value of 1 and explored cohort-specific values of sensitivity and specificity. All diabetes risk models were adjusted by standard risk factors, also included in the calibration equations. Additionally, we stratified our discrete proportional hazards models on age in 10-year categories and DM-C or OS membership to better approximate previous analyses. Because BMI may be only a mediator for energy intake or may possibly also be an independent risk factor, it is not clear whether adjusting for BMI in our diabetes risk model is appropriate due to the challenge of overcontrolled or undercontrolled models, as discussed in Tinker et al. \cite{tinker2011biomarker} Thus, we ran each outcome model with and without BMI.

To fit the naive model, we used the binomial generalized linear model with the complementary log-log link. To fit the model corrected for covariate error only, we used this same approach, then adopted the post-hoc matrix correction and corresponding variance adjustment described in the body of this paper. We applied our proposed approach to correct for error in both the self-reported diabetes outcome and dietary exposures. In all models, we used log values of dietary energy, protein, and protein density. We present hazard ratios (HR) and 95\% confidence intervals (CI) associated with a 20\% increase in consumption.

\subsection{Results}

Incident diabetes was reported in 3053 (4.7\%) of the 65,358 participants of analytic cohort. Table \ref{tableHR} shows the results for the three different analysis approaches. In the BMI-adjusted analysis, the HR (95\% CI) for a 20\% increase in energy intake was 0.822 (0.512, 1.318) for the proposed approach compared to 1.041 (0.758, 1.492) for the covariate-error adjusted method and 1.002 (0.986, 1.018) for the naive approach. Note, however, that the incident diabetes is not significantly associated with increasing energy in any of these three models. Without BMI in the outcome model, the proposed method estimated a HR of 1.189 (0.836, 1.692) for a 20\% increase in energy intake, compared to 1.421 (1.043, 1.938) for the covariate-error adjusted method and 1.024 (1.008, 1.040) for the naive method. In this case, adjusting for error in the self-reported outcome led to qualitatively different results in that the HR was about 20\% smaller and no longer significant.

When we apply the proposed method, a 20\% increase in protein intake is associated with a 1.077 (0.978, 1.186) HR, compared to a HR of 1.121 (1.036, 1.213) for the covariate-error adjusted method and 1.024 (1.010, 1.039) for the naive approach.  When we do not adjust for BMI, all three approaches result in HRs that are significantly associated with an increase in protein consumption. For protein density, whether or not we adjust for BMI, all three approaches show that a 20\% increase in intake is positively associated with risk of diabetes. When we adjust for BMI, the HR estimated by the proposed method, 1.266 (1.115, 1.436), is fairly similar to the HR estimated by the method that adjusts for covariate error only, 1.243 (1.125, 1.374), and somewhat higher than the HR estimated by the naive method, 1.100 (1.064, 1.137).  We note some of our HRs differ from the results reported by Tinker et al. \cite{tinker2011biomarker} We believe this is due to a few discrepancies in the analytical dataset and model and is discussed further in Section S4 of the Supplementary Materials.

%For example, adjusting for covariate error, Tinker et al. (2011) reports a HR (95\% CI) of 2.41 (2.06, 2.82) for a 20\% increase in energy intake in the outcome model that omitted BMI, compared to our 1.421 (1.043, 1.938).

In Table \ref{suppWHI} of the Supplementary Materials, we present a WHI data analysis results table that ignores the issue of an imperfect baseline self-report and assumes the negative predictive value is 1. For energy and protein density, assuming baseline self-reports are perfect does not qualitatively change our results. However, for protein, the HR (95\% confidence interval) estimated by the proposed method is 1.077 (0.978, 1.186) when the negative predictive value is set to 0.96, but changes to 1.107 (1.025, 1.195) when the negative predictive value is set to 1. Here, we see that because our estimate is so close to a boundary, incorporating the uncertainty at baseline into our analyses does slightly change our results. 

Since we analyzed data on participants from two different cohorts, the WHI DM-C trial and the WHI OS, we investigated how cohort-specific sensitivity and specificity might impact our HR estimates.  We used a weighted-average approach to select sensitivity and specificity values for the DM-C and OS trials such that the overall values worked out to be 0.61 and 0.995, respectively. One might hypothesize that the clinical trial (WHI DM-C) recorded data with higher accuracy than the larger observational study (OS), though in our analysis we also consider the possibility that sensitivity and specificity are higher for the observational study. Table \ref{supp1} in the Supplementary Materials presents the results of this analysis. We observe that implementing slightly variable cohort-specific sensitivity and specificity values was not enough to qualitatively impact our conclusions regarding the significance of the association between an increase in intake of dietary energy, protein, or protein density with the risk of diabetes.

\section{Discussion}\label{discussion}
In settings such as large epidemiological studies, where outcomes or complex exposures are often collected by self-report, both the exposure and outcome of interest can be subject to measurement error. This was observed in our data example from the WHI, but has also been observed in other cohorts where data were reliant on routinely collected electronic health records data.\citep{shepherd2011accounting, oh2019raking} This paper presents a method to accommodate errors in continuous covariates and a discrete failure time outcome variable when sensitivity and specificity of the error-prone outcome are known; when error rates are unknown, our method can be used as a sensitivity analysis  using hypothesized values. The proposed method can be applied when, for a subset, there is either a gold standard measure of the exposure or a second measure with independent, unbiased (classical) measurement error available. For the WHI, the calibration subset containing the variable with classical measurement error was sampled after baseline with the assumption that the measurement error model did not change over time. 

We studied the relative performance of the proposed method under various settings of sensitivity, specificity, error variance of the exposure, and censoring rate, including those where ignoring the measurement error led to extreme bias in the regression parameters of interest. In all settings studied, our method led to nearly unbiased estimates of the regression parameters, maintaining bias of less than approximately 19\% for non-zero regression parameters and generally much less bias when the underlying log-hazard parameter $\beta$ was of moderate size (e.g., $\log(1.5)$). Furthermore, our variance estimator performed favorably, as evidenced by the coverage probability and ASEs that closely resembled ESEs. Our variance estimator assumes approximate independence of $\hat{\beta}^*$ and $\hat{\delta}$. While we have not verified independence of these components for all settings, even in our settings where the calibration subset was 50\% of the cohort, we observed no appreciable correlation between these estimates (data not shown). If there is concern that this approximate independence does not hold, one could instead consider a bootstrap approach for variance estimation. For our simulations where $\beta_{X1}=0$, we observed that type I error rates were preserved. Our adjustment for covariate error relied on a regression calibration type adjustment. As expected from previous literature, this method performs best when the regression parameter corresponding to the error-prone covariate is of modest size, the error in the covariate is normally distributed, and the censoring rate is high (i.e. the event of interest is rare). Our method in particular shows more appreciable bias when the regression parameter is of large size, e.g. $\beta_{X1}=\log(3)$, especially for a lower censoring rate. This method proved to be fairly robust to changes in the distribution of the error in $X$ studied; for more extreme deviations from normality, this may no longer be true. Our method also performs favorably after stratifying on one or more covariates. Lastly, the proposed method works well under simulation parameters that mimic the structure of the WHI data. In all scenarios explored, the proposed method substantially outperformed the naive method, which repeatedly showed severe bias and minimal coverage. For settings different from those studied, one might consider conducting additional numerical studies. 

The method introduced in this paper is applied to data from 65,358 post-menopausal women enrolled in the WHI to assess the association between energy, protein, and protein density  intake and the risk of incident diabetes, adjusting for error in self-reported exposures and outcome. Hazard ratios obtained for all exposures were considerably different than those from the naive analyses ignoring the error in both diabetes status and dietary intake and those that only adjusted for error in dietary intake. In some cases, our proposed method led to qualitatively different conclusions in that the parameter of interest was no longer statistically significant. For the case of non-differential outcome error, this stems largely from the increased uncertainty in the results coming from the uncertain outcomes. These conclusions demonstrate the importance of adjusting for errors in both outcomes and covariates.  

Our proposed method offers a practical approach to estimating the association between a covariate and a discrete time-to-event outcome, when both are recorded with error. A limitation of our approach stems from the curse of dimensionality that can accompany discrete data in settings where the visit times are irregular, which can cause the number of parameters to grow with the number of subjects in the data. It is impractical to assume that in a real data setting, all subjects' visit times fall on the same schedule in the study (e.g. exactly annually). Thus, we must make a compromise depending on how many parameters the data can stably support. Ultimately, the data should help inform a reasonable decision regarding the number of intervals to consider for analyses of this type. Sensitivity analyses can be also be conducted to examine whether the number or choice of discrete time intervals affected study estimates.  
In many cohort studies with long-term follow-up like the WHI, there is a specified visit schedule in the study protocol. If all subjects adhere to this schedule with little variation, this naturally leads to the discrete-time framework with a common set of possible visit times across all individuals. Frequently in these studies, including the WHI, the observed visit schedule varies across subjects.  To apply the proposed method in our WHI example, we made some simplifying assumptions.  Since our analytical data set included only the amount of time that elapsed between enrollment and the first occurrence of self-reported diabetes or censoring time recorded on a continuous timescale, we rounded the censored event-time to the nearest annual visit date and assumed the outcome or censoring event occurred sometime between that visit and the prior annual visit. If data are available on the timing of all visits, the likelihood could be adapted to allow for longer intervals between visits for some individuals (i.e. missed visits). 

We note that for the case of self-reported data, we assume that each subject is followed up until the first positive, as it is not expected that a new diagnosis would be subsequently recorded. This assumption corresponds to the applied setting in which self-reported disease incidence stops after the first positive report. However, the model by \citet{gu2015semiparametric} and thus the proposed approach do allow for a more flexible framework and can accommodate repeated testing. As an example, this approach can be applied to a data set containing repeat blood test results, such as those used in monitoring for cancer relapse.

A potential limitation of our work is the reliance of the proposed method on the assumption that given the true disease status at each visit, the error-prone outcomes are independent. 
%We recognize that this assumption may not always be realistic, particularly for the case of self-reported data. Nonetheless, our method could be applied to other settings where the error-prone outcome of interest is not self-reported. Consider, for example, annual mammograms for breast cancer screening, where we can assume that different radiologists are interpreting the mammograms at each annual visit. The annual mammogram results may be error-prone due to rater error or biological variability. In this example, it is reasonable to assume that the mammogram results are independent given the true time of development of breast cancer. In future work, one might consider relaxing this assumption and positing a different error model for the outcome of interest.
In the WHI data, we assume the self-reported outcomes are far enough apart that there are a number of random processes affecting a subject's knowledge and interpretation of the outcomes questionnaire that make this independence assumption reasonable; however, this assumption may not always be realistic, particularly for the case of self-reported data. We note that our method is applicable more generally to settings where the error-prone outcome of interest is not self-reported, but derived say from an objective biomarker for which this assumption may be more reasonable. 
%Consider, for example, annual mammograms for breast cancer screening, where we can assume that different radiologists are interpreting the mammograms at each annual visit. The annual mammogram results may be error-prone due to rater error or biological variability. In this example, it is reasonable to assume that the mammogram results are independent given the true time of development of breast cancer. 
In future work, one might consider a similar framework to the one proposed which relaxes this assumption by positing a more complex error model for the outcome of interest, such as one with sensitivity and specificity potentially dependent on covariates or previous responses.

The increasing reliance of clinical research on self-administered questionnaires or administrative databases in epidemiological studies has led to more attention being given to methods to correct for measurement error.  \citet{gu2015semiparametric} conducted a sensitivity analysis to show how changes in sensitivity, specificity, and negative predictive shifted the estimated hazard ratio of statin use on the risk of incident diabetes in data from the WHI. The results showed that the estimated hazard ratio is highly sensitive to changes in specificity and modestly sensitive to changes in sensitivity and negative predictive value. This analysis helps illustrate the importance of having accurate values of sensitivity and specificity in the proposed method. Our sensitivity analysis showed that while varying sensitivity and specificity by cohort did not qualitatively change the results in our particular example,  the hazard ratio estimates are much more vulnerable to changes in specificity when the event of interest is as rare as it is in the WHI data (diabetes incidence = 4.7\%). Thus, we emphasize the importance of employing correct values of sensitivity and specificity, especially when they might vary by some demographic factor or group membership.  

This paper explored the incorporation of the negative predictive value into the analyses to handle  misclassification at baseline. Evidence suggests that some women in the WHI who provided a negative self-report of diabetes at baseline were actually diabetic. A question of interest is whether mistakenly excluding women who were false positives can induce bias. It has been previously reported that when all potential confounders are adjusted for in the outcome model and the missing at random (MAR) assumption is satisfied, missing data should not cause bias.\citep{groenwold2011dealing} Furthermore, given that positive predictive value is assumed to be quite high in the motivating data example, we did not explore the issue further in this paper. This exclusion criteria-related matter may be more relevant in other cohorts, particularly if the reason for exclusion is related to some unobserved characteristic. 

A worthwhile extension of this work might consider incorporating covariate-specific or even subject-specific sensitivity and specificity, particularly when these values are no longer assumed to be known constants and need to be estimated along with the outcome model parameters. Such an extension would require a validation or calibration subset to also contain information on the measurement error structure of the self-reported outcome. When the outcome is rare, such a cohort can be difficult to construct prospectively as validation subsets are generally of fairly modest size due to cost. Efficient choices of a validation sampling design and development of analysis methods that provide consistent estimates of the target parameter are two important areas of future research.

%  The \backmatter command formats the subsequent headings so that they
%  are in the journal style.  Please keep this command in your document
%  in this position, right after the final section of the main part of 
%  the paper and right before the Acknowledgements, Supplementary Materials,
%  and References sections. 

%  This section is optional.  Here is where you will want to cite
%  grants, people who helped with the paper, etc.  But keep it short!

\section*{Acknowledgements}
This work was supported in part by NIH grant R01-AI131771 and PCORI grant R-1609-36207. The authors would like to thank the investigators of the Women’s Health Initiative (WHI) for the use of their data. A short list of WHI investigators can be found here: \href{https://www.whi.org/researchers/Documents%20%20Write%20a%20Paper/WHI%20Investigator%20Short%20List.pdf}{https://www.whi.org/researchers}.
 The WHI program is funded by the National Heart, Lung, and Blood Institute, National Institutes of Health, U.S. Department of Health and Human Services through contracts HHSN268201600018C, HHSN268201600001C, \\ HHSN268201600002C, HHSN268201600003C, and HHSN268201600004C.

\section*{Supporting Information}

R code implementing all simulations and a sample data analysis illustrating our method with simulated data is available on GitHub at \href{https://github.com/lboe23/Outcome-Error-RC}{https://github.com/lboe23/Outcome-Error-RC}. Additionally, R code demonstrating how to apply the proposed method to a simulated data set is presented in Section S1 of the Supplementary Materials.

\section*{Data Availability Statement}

The data used in this paper can be obtained through submission and approval of a manuscript proposal to the Women's Health Initiative Publications and Presentations Committee, as described on the WHI website.\citep{WHIdatacite} For more details, see \href{https://www.whi.org/researchers/data/Pages/Home.aspx}{https://www.whi.org/researchers/data}.

\vspace*{-8pt}

\bibliography{bibliofinal}

\begin{table}
\centering
\begin{adjustbox}{totalheight=\textheight}
      \begin{threeparttable}[t]
            \caption{The mean percent (\%) biases, average standard errors (ASE), empirical standard errors (ESE) and coverage probabilities (CP) are given for 1000 simulated data sets for the proposed method and naive method with $\beta_{X1}=\log(1.5)$, $\beta_{Z1}=\log(0.7)$, and $\beta_{Z2}=\log(1.3)$; $e$ is normally distributed with mean zero.}\label{table1sims}
   
   \begin{tabular}{ccccccccccc}
\hline
\multicolumn{3}{l}{ $Se\tnote{1} =0.80, Sp\tnote{2} =0.90$} & \multicolumn{4}{c}{Proposed}  & \multicolumn{4}{c}{Naive} \\ \cmidrule(r){4-7} \cmidrule{8-11} $\hat{\delta}_{(1)}\tnote{3}$ & CR\tnote{4} & $\beta$ & \% Bias & ASE & ESE & CP & \% Bias & ASE & ESE & \multicolumn{1}{c}{CP} \\ 

\hline
0.60 & 0.90 & $\beta_{X1}$  & $\phantom{-}1.616$ & 0.200 & 0.204 & 0.950 & $-88.03$ & 0.046 & 0.046 & 0.000 \\ 
 &  & $\beta_{Z1}$  &  $-1.094$ & 0.143 & 0.142 & 0.945 & $-79.22$ & 0.057 & 0.058 & 0.002 \\ 
 &  & $\beta_{Z2}$  &  $-3.731$ & 0.143 & 0.143 & 0.945 & $-84.07$ & 0.057 & 0.054 & 0.021 \\ 
 & 0.55 & $\beta_{X1}$  &   $-1.231$ & 0.093 & 0.094 & 0.949 & $-68.11$ & 0.038 & 0.038 & 0.000 \\ 
 &  & $\beta_{Z1}$  & $-1.055$ & 0.067 & 0.066 & 0.958 & $-43.46$ & 0.047 & 0.046 & 0.079 \\ 
 &  & $\beta_{Z2}$  & $-3.018$ & 0.066 & 0.065 & 0.957 & $-53.48$ & 0.046 & 0.045 & 0.133 \\ 
0.30 & 0.90 & $\beta_{X1}$  &  $\phantom{-}1.840$ & 0.283 & 0.286 & 0.954 & $-93.88$ & 0.033 & 0.033 & 0.000 \\ 
 &  & $\beta_{Z1}$  & $-1.233$ & 0.151 & 0.151 & 0.947 & $-82.46$ & 0.054 & 0.055 & 0.001 \\ 
 &  & $\beta_{Z2}$  & $-4.212$ & 0.151 & 0.150 & 0.945 & $-79.74$ & 0.054 & 0.052 & 0.025 \\ 
 & 0.55 & $\beta_{X1}$  & $-2.246$ & 0.131 & 0.133 & 0.940 & $-84.02$ & 0.027 & 0.027 & 0.000 \\ 
 &  & $\beta_{Z1}$  &  $-1.967$ & 0.071 & 0.069 & 0.951 & $-52.48$ & 0.045 & 0.044 & 0.008 \\ 
 &  & $\beta_{Z2}$  &  $-3.899$ & 0.070 & 0.068 & 0.956 & $-42.08$ & 0.045 & 0.044 & 0.306 \\ 
 \hline \hline

\multicolumn{3}{l}{ $Se=0.90, Sp=0.80$} & \multicolumn{4}{c}{Proposed}  & \multicolumn{4}{c}{Naive} \\ \cmidrule(r){4-7} \cmidrule{8-11} $\hat{\delta}_{(1)}$ & CR & $\beta$ & \% Bias & ASE & ESE & CP & \% Bias & ASE & ESE & \multicolumn{1}{c}{CP} \\ 
\hline 
0.60 & 0.90 & $\beta_{X1}$  &  $\phantom{-}0.391$ & 0.210 & 0.209 & 0.957 & $-93.08$ & 0.037 & 0.037 & 0.000 \\ 
 &  & $\beta_{Z1}$  &  $-3.692$ & 0.150 & 0.153 & 0.942 & $-91.96$ & 0.046 & 0.045 & 0.001 \\ 
 & 0.55 & $\beta_{X1}$  &  $-1.246$ & 0.094 & 0.093 & 0.960 & $-77.95$ & 0.034 & 0.035 & 0.000 \\ 
 &  & $\beta_{Z1}$  &  $-1.188$ & 0.068 & 0.067 & 0.951 & $-61.05$ & 0.042 & 0.042 & 0.001 \\ 
 &  & $\beta_{Z2}$  & $-3.502$ & 0.067 & 0.066 & 0.953 & $-68.46$ & 0.042 & 0.042 & 0.014 \\ 
0.30 & 0.90 & $\beta_{X1}$  & $\phantom{-}0.665$ & 0.296 & 0.291 & 0.967 & $-96.33$ & 0.026 & 0.026 & 0.000 \\ 
 &  & $\beta_{Z1}$  &  $-0.963$ & 0.158 & 0.160 & 0.951 & $-90.21$ & 0.044 & 0.044 & 0.000 \\ 
 &  & $\beta_{Z2}$  &  $-4.214$ & 0.158 & 0.160 & 0.947 & $-89.56$ & 0.044 & 0.043 & 0.001 \\ 
 & 0.55 & $\beta_{X1}$  &  $-2.034$ & 0.133 & 0.130 & 0.964 & $-88.87$ & 0.024 & 0.024 & 0.000 \\ 
 &  & $\beta_{Z1}$  &  $-1.994$ & 0.072 & 0.070 & 0.950 & $-67.22$ & 0.040 & 0.040 & 0.000 \\ 
 &  & $\beta_{Z2}$  & $-4.420$ & 0.071 & 0.069 & 0.959 & $-60.63$ & 0.040 & 0.040 & 0.029 \\
\hline
\multicolumn{3}{l}{ $Se=1, Sp=1$} & \multicolumn{4}{c}{Truth}  \\ \cmidrule(r){4-7}   & CR & $\beta$ & \% Bias & ASE & ESE & CP \\
   \hline
  & 0.90 & $\beta_{X1}$ & $\phantom{-}0.163$ & 0.108 & 0.109 & 0.944 \\ 
  & & $\beta_{Z1}$ &   $\phantom{-}0.205$ & 0.107 & 0.107 & 0.953 \\ 

  &   &$\beta_{Z2}$ & $-0.586$ & 0.107 & 0.109 & 0.949 \\ 
  & 0.55 &  $\beta_{X1}$  & $\phantom{-}0.639$ & 0.052 & 0.052 & 0.948 \\ 
  & &  $\beta_{Z1}$ &  $\phantom{-}0.345$ & 0.052 & 0.051 & 0.949 \\ 
  &  &  $\beta_{Z2}$ & $-0.383$ & 0.052 & 0.052 & 0.952 \\ 
   \hline
\end{tabular}
 \begin{tablenotes}
     \item[1] $Se=$ Sensitivity
     \item[2] $Sp=$ Specificity    
     \item[3] $\hat{\delta}_{(1)}=$ estimate of attenuation coefficient
     \item[4] $CR=$ True censoring rate 
   \end{tablenotes}
    \end{threeparttable}
    \end{adjustbox}
\end{table}

\begin{table}   
\centering
\begin{adjustbox}{totalheight=\textheight}
          \begin{threeparttable}[t]
            \caption{The mean percent (\%) biases, average standard errors (ASE), empirical standard errors (ESE) and coverage probabilities (CP) are given for 1000 simulated data sets for the proposed method and naive method with $\beta_{X1}=\log(3)$, $\beta_{Z1}=\log(0.7)$, and $\beta_{Z2}=\log(1.3)$; $e$ is normally distributed with mean zero.}
    \label{table2sims}
\begin{tabular}{ccccccccccc}
 \hline
\multicolumn{3}{l}{ $Se\tnote{1} =0.80, Sp\tnote{2} =0.90$} & \multicolumn{4}{c}{Proposed}  & \multicolumn{4}{c}{Naive} \\ \cmidrule(r){4-7} \cmidrule{8-11} $\hat{\delta}_{(1)}\tnote{3}$ & CR\tnote{4} & $\beta$ & \% Bias & ASE & ESE & CP & \% Bias & ASE & ESE & \multicolumn{1}{c}{CP} \\ 
\hline
0.60 & 0.90 & $\beta_{X1}$  & $-3.442$ & 0.211 & 0.213 & 0.946 & $-88.61$ & 0.047 & 0.048 & 0.000 \\ 
 &  & $\beta_{Z1}$  &  $-6.773$ & 0.146 & 0.145 & 0.941 & $-78.03$ & 0.057 & 0.059 & 0.002 \\ 
 &  & $\beta_{Z2}$  &  $-9.280$ & 0.145 & 0.142 & 0.948 & $-88.07$ & 0.057 & 0.054 & 0.012 \\ 
 & 0.55 & $\beta_{X1}$  & $-12.71$ & 0.111 & 0.101 & 0.752 & $-72.45$ & 0.040 & 0.038 & 0.000 \\ 
 &  & $\beta_{Z1}$  &  $-12.57$ & 0.075 & 0.068 & 0.916 & $-45.77$ & 0.047 & 0.047 & 0.078 \\ 
 &  & $\beta_{Z2}$  &  $-14.42$ & 0.075 & 0.066 & 0.952 & $-67.90$ & 0.047 & 0.045 & 0.032 \\ 
0.30 & 0.90 & $\beta_{X1}$  & $-4.532$ & 0.296 & 0.295 & 0.951 & $-94.26$ & 0.033 & 0.033 & 0.000 \\ 
 &  & $\beta_{Z1}$  &  $-8.063$ & 0.156 & 0.152 & 0.944 & $-86.52$ & 0.054 & 0.056 & 0.000 \\ 
 &  & $\beta_{Z2}$  & $-11.64$ & 0.155 & 0.149 & 0.951 & $-77.03$ & 0.054 & 0.052 & 0.025 \\ 
  
 & 0.55 & $\beta_{X1}$  & $-16.88$ & 0.154 & 0.137 & 0.766 & $-86.56$ & 0.028 & 0.027 & 0.000 \\ 
 &  & $\beta_{Z1}$  &  $-16.75$ & 0.080 & 0.071 & 0.899 & $-67.26$ & 0.045 & 0.045 & 0.000 \\ 
 
 &  & $\beta_{Z2}$  &  $-18.69$ & 0.080 & 0.069 & 0.956 & $-42.46$ & 0.045 & 0.044 & 0.299 \\
  \hline
\multicolumn{3}{l}{ $Se=0.90, Sp=0.80$} & \multicolumn{4}{c}{Proposed}  & \multicolumn{4}{c}{Naive} \\ \cmidrule(r){4-7} \cmidrule{8-11} $\hat{\delta}_{(1)}$ & CR & $\beta$ & \% Bias & ASE & ESE & CP & \% Bias & ASE & ESE & \multicolumn{1}{c}{CP} \\ 
\hline
0.60 & 0.90 & $\beta_{X1}$  &  $-3.581$ & 0.220 & 0.221 & 0.945 & $-93.60$ & 0.038 & 0.039 & 0.000 \\ 
 &  & $\beta_{Z1}$  & $-6.164$ & 0.152 & 0.153 & 0.936 & $-87.76$ & 0.046 & 0.047 & 0.000 \\ 
 &  & $\beta_{Z2}$  &  $-8.663$ & 0.151 & 0.150 & 0.954 & $-94.13$ & 0.046 & 0.045 & 0.000 \\ 
 & 0.55 & $\beta_{X1}$  & $-12.65$ & 0.112 & 0.103 & 0.764 & $-80.88$ & 0.035 & 0.035 & 0.000 \\ 
 &  & $\beta_{Z1}$  &  $-12.64$ & 0.076 & 0.071 & 0.915 & $-62.26$ & 0.042 & 0.043 & 0.001 \\ 
 &  & $\beta_{Z2}$  &  $-14.65$ & 0.076 & 0.068 & 0.939 & $-78.22$ & 0.042 & 0.042 & 0.001 \\ 
0.30 & 0.90 & $\beta_{X1}$  &  $-4.585$ & 0.309 & 0.298 & 0.963 & $-96.73$ & 0.027 & 0.027 & 0.000 \\ 
 &  & $\beta_{Z1}$  &  $-7.171$ & 0.163 & 0.159 & 0.947 & $-92.46$ & 0.044 & 0.045 & 0.000 \\ 
 &  & $\beta_{Z2}$  &  $-11.06$ & 0.162 & 0.157 & 0.954 & $-88.01$ & 0.044 & 0.043 & 0.000 \\ 
 & 0.55 & $\beta_{X1}$  &  $-16.67$ & 0.156 & 0.138 & 0.772 & $-90.62$ & 0.025 & 0.024 & 0.000 \\ 
 &  & $\beta_{Z1}$  &  $-16.65$ & 0.082 & 0.073 & 0.901 & $-77.08$ & 0.040 & 0.041 & 0.000 \\ 
 &  & $\beta_{Z2}$  &  $-18.86$ & 0.081 & 0.070 & 0.943 & $-60.56$ & 0.040 & 0.040 & 0.025 \\ 
\hline
\multicolumn{3}{l}{ $Se=1, Sp=1$} & \multicolumn{4}{c}{Truth}  \\ \cmidrule(r){4-7}   & CR & $\beta$ & \% Bias & ASE & ESE & CP \\
   \hline 
   & 0.90 & $\beta_{X1}$ &  $\phantom{-}0.565$ & 0.115 & 0.116 & 0.951 \\ 
   &  & $\beta_{Z1}$ & $-0.222$ & 0.108 & 0.108 & 0.949 \\ 
   &  & $\beta_{Z2}$ &   $-0.347$ & 0.108 & 0.110 & 0.948 \\ 
   & 0.55 & $\beta_{X1}$ &   $\phantom{-}0.605$ & 0.063 & 0.064 & 0.944 \\ 
   &  & $\beta_{Z1}$ &  $\phantom{-}0.264$ & 0.054 & 0.054 & 0.952 \\ 
   &  & $\beta_{Z2}$ & $-0.162$ & 0.054 & 0.052 & 0.955 \\ 
   \hline
  \end{tabular}
 \begin{tablenotes}
     \item[1] $Se=$ Sensitivity
     \item[2] $Sp=$ Specificity    
     \item[3] $\hat{\delta}_{(1)}=$ estimate of attenuation coefficient
     \item[4] $CR=$ True censoring rate
   \end{tablenotes}
    \end{threeparttable}
    \end{adjustbox}
\end{table}

\begin{table} 
    \centering
    \begin{adjustbox}{totalheight=\textheight}
      \begin{threeparttable}[t]
            \caption{The mean percent (\%) biases, average standard errors (ASE), empirical standard errors (ESE) and coverage probabilities (CP) are given for 1000 simulated data sets for the proposed method and naive method with $\beta_{X1}=\log(1.5)$, $\beta_{Z1}=\log(0.7)$, and $\beta_{Z2}=\log(1.3)$; $e$ is distributed as either a \textit{t} with 4 df or as $.4\mathcal{N}(0, 1) + .6\mathcal{N}(2, 1.5)$.}\label{table3sims}
   \begin{tabular}{ccccccccccc}
\hline
\multicolumn{3}{l}{ $Se\tnote{1} =0.80, Sp\tnote{2} =0.90$} & \multicolumn{4}{c}{Proposed}  & \multicolumn{4}{c}{Naive} \\ \cmidrule(r){4-7} \cmidrule{8-11} $e$ \tnote{3} & CR\tnote{4} & $\beta$ & \% Bias & ASE & ESE & CP & \% Bias & ASE & ESE & \multicolumn{1}{c}{CP} \\ 

\hline
\textit{t}\tnote{5} & 0.90 & $\beta_{X1}$  & $-2.238$ & 0.291 & 0.300 & 0.953 & $-94.50$ & 0.031 & 0.031 & 0.000 \\ 
 &  & $\beta_{Z1}$  & $\phantom{-}0.622$ & 0.152 & 0.157 & 0.951 & $-81.80$ & 0.054 & 0.054 & 0.000 \\ 
 &  & $\beta_{Z2}$  & $\phantom{-}2.529$ & 0.152 & 0.148 & 0.957 & $-76.70$ & 0.054 & 0.053 & 0.033 \\ 
 & 0.55 & $\beta_{X1}$  &  $-0.646$ & 0.140 & 0.153 & 0.940 & $-85.54$ & 0.025 & 0.026 & 0.000 \\ 
 &  & $\beta_{Z1}$  & $-2.362$ & 0.072 & 0.072 & 0.950 & $-53.37$ & 0.045 & 0.044 & 0.013 \\ 
 &  & $\beta_{Z2}$  &  $-3.303$ & 0.071 & 0.072 & 0.950 & $-39.35$ & 0.044 & 0.044 & 0.335 \\ 
$mix$\tnote{6} & 0.90 & $\beta_{X1}$  & $\phantom{-}0.394$ & 0.335 & 0.330 & 0.955 & $-95.64$ & 0.027 & 0.028 & 0.000 \\ 
 &  & $\beta_{Z1}$  &  $-1.015$ & 0.158 & 0.158 & 0.953 & $-83.29$ & 0.054 & 0.055 & 0.000 \\ 
 &  & $\beta_{Z2}$  & $-0.800$ & 0.156 & 0.152 & 0.962 & $-75.25$ & 0.054 & 0.056 & 0.055 \\ 
 & 0.55 & $\beta_{X1}$  & $-1.081$ & 0.156 & 0.151 & 0.958 & $-88.74$ & 0.022 & 0.022 & 0.000 \\ 
 &  & $\beta_{Z1}$  &  $-2.415$ & 0.074 & 0.070 & 0.958 & $-55.28$ & 0.044 & 0.045 & 0.010 \\ 
 &  & $\beta_{Z2}$  & $-2.083$ & 0.073 & 0.070 & 0.964 & $-36.40$ & 0.044 & 0.045 & 0.419 \\
 \hline
\multicolumn{3}{l}{ $Se=0.90, Sp=0.80$} & \multicolumn{4}{c}{Proposed}  & \multicolumn{4}{c}{Naive} \\ \cmidrule(r){4-7} \cmidrule{8-11} $\hat{\delta}_{(1)}$ & CR & $\beta$ & \% Bias & ASE & ESE & CP & \% Bias & ASE & ESE & \multicolumn{1}{c}{CP} \\ 

\hline
\textit{t} & 0.90 & $\beta_{X1}$  &  $-3.792$ & 0.305 & 0.316 & 0.942 & $-96.91$ & 0.025 & 0.025 & 0.000 \\ 
 &  & $\beta_{Z1}$  &  $\phantom{-}1.848$ & 0.160 & 0.165 & 0.948 & $-89.40$ & 0.044 & 0.044 & 0.000 \\ 
 
 &  & $\beta_{Z2}$  &  $\phantom{-}3.386$ & 0.159 & 0.158 & 0.959 & $-86.97$ & 0.044 & 0.044 & 0.000 \\ 
 
 & 0.55 & $\beta_{X1}$  &  $-1.119$ & 0.141 & 0.159 & 0.933 & $-90.02$ & 0.023 & 0.024 & 0.000 \\ 

 &  & $\beta_{Z1}$  & $-1.666$ & 0.073 & 0.073 & 0.940 & $-67.86$ & 0.040 & 0.040 & 0.000 \\ 
 &  & $\beta_{Z2}$  &  $-3.048$ & 0.072 & 0.074 & 0.944 & $-58.35$ & 0.040 & 0.040 & 0.031 \\ 
 
$mix$ & 0.90 & $\beta_{X1}$  &  $-0.975$ & 0.350 & 0.346 & 0.952 & $-97.65$ & 0.022 & 0.024 & 0.000 \\ 
 
 &  & $\beta_{Z1}$  &  $-1.354$ & 0.166 & 0.162 & 0.961 & $-90.94$ & 0.043 & 0.044 & 0.000 \\ 
 
 &  & $\beta_{Z2}$  &  $\phantom{-}0.585$ & 0.164 & 0.160 & 0.955 & $-86.57$ & 0.043 & 0.045 & 0.000 \\ 
 
 & 0.55 & $\beta_{X1}$  &  $-1.904$ & 0.159 & 0.155 & 0.955 & $-92.26$ & 0.020 & 0.021 & 0.000 \\ 
  
 &  & $\beta_{Z1}$  & $-2.590$ & 0.075 & 0.072 & 0.954 & $-69.29$ & 0.040 & 0.040 & 0.000 \\ 
  
 &  & $\beta_{Z2}$  & $-1.350$ & 0.074 & 0.069 & 0.967 & $-56.67$ & 0.040 & 0.039 & 0.036 \\
\hline
\multicolumn{3}{l}{ $Se=1, Sp=1$} & \multicolumn{4}{c}{Truth}  \\ \cmidrule(r){4-7}   & CR & $\beta$ & \% Bias & ASE & ESE & CP \\
   \hline
&  0.90 & $\beta_{X1}$ & $\phantom{-}0.002$ & 0.108 & 0.106 & 0.959 \\ 
& & $\beta_{Z1}$ & $\phantom{-}0.034$ & 0.108 & 0.109 & 0.951 \\ 
&   & $\beta_{Z2}$ &  $\phantom{-}1.032$ & 0.107 & 0.106 & 0.961 \\ 
 &0.55 & $\beta_{X1}$  & $\phantom{-}0.395$ & 0.053 & 0.052 & 0.952 \\ 
& & $\beta_{Z1}$  & $-0.462$ & 0.052 & 0.052 & 0.948 \\ 
&  & $\beta_{Z2}$ &  $-0.300$ & 0.052 & 0.050 & 0.954 \\ 
   \hline
\end{tabular}
 \begin{tablenotes}
     \item[1] $Se=$ Sensitivity
     \item[2] $Sp=$ Specificity    
     \item[3] $e$ refers to the distribution of the error
     \item[4] $CR=$ True censoring rate
     \item[5] \textit{t} with 4 df
     \item[6] Mixture of two normals, i.e. $.4\mathcal{N}(0, 1) + .6\mathcal{N}(2, 1.5)$
   \end{tablenotes}
    \end{threeparttable}
    \end{adjustbox}
\end{table}

\begin{table}
    \centering
    \begin{adjustbox}{totalheight=\textheight}
                  \begin{threeparttable}[t]
    \caption{The mean percent (\%) biases, average standard errors (ASE), empirical standard errors (ESE) and coverage probabilities (CP) are given for 1000 simulated data sets for the proposed method and naive method, when both allow for strata-specific baseline hazards. We assume four equally-sized strata. Let $\beta_{X1}=\log(1.5)$, $\beta_{Z1}=\log(0.7)$, and $\beta_{Z2}=\log(1.3)$; $e$ is normally distributed with mean zero.}
    \label{table4sims}
    \begin{tabular}{ccccccccccc}
 \hline
\multicolumn{3}{l}{ $Se\tnote{1} =0.80, Sp\tnote{2} =0.90$} & \multicolumn{4}{c}{Proposed}  & \multicolumn{4}{c}{Naive} \\ \cmidrule(r){4-7} \cmidrule{8-11} $\hat{\delta}_{(1)}\tnote{3}$ & CR\tnote{4} & $\beta$ & \% Bias & ASE & ESE & CP & \% Bias & ASE & ESE & \multicolumn{1}{c}{CP} \\ 

\hline
0.60 & 0.90 & $\beta_{X1}$  & $\phantom{-}1.893$ & 0.202 & 0.199 & 0.961 & $-88.44$ & 0.047 & 0.046 & 0.000 \\ 
   &  & $\beta_{Z1}$  &  $\phantom{-}3.249$ & 0.145 & 0.148 & 0.954 & $-78.36$ & 0.057 & 0.058 & 0.001 \\ 
 
 &  & $\beta_{Z2}$  &   $-0.263$ & 0.144 & 0.151 & 0.946 & $-81.41$ & 0.057 & 0.058 & 0.044 \\ 

 & 0.55 & $\beta_{X1}$  &   $-0.489$ & 0.094 & 0.089 & 0.965 & $-68.31$ & 0.038 & 0.038 & 0.000 \\ 
  
 &  & $\beta_{Z1}$  &  $\phantom{-}0.001$ & 0.068 & 0.066 & 0.960 & $-42.99$ & 0.047 & 0.047 & 0.095 \\ 
  
 &  & $\beta_{Z2}$  & $-0.885$ & 0.067 & 0.066 & 0.958 & $-52.09$ & 0.047 & 0.048 & 0.172 \\ 
  
0.30 & 0.90 & $\beta_{X1}$  & $\phantom{-}1.036$ & 0.286 & 0.280 & 0.959 & $-94.20$ & 0.033 & 0.033 & 0.000 \\ 
  
 &  & $\beta_{Z1}$  &  $\phantom{-}2.777$ & 0.153 & 0.154 & 0.956 & $-81.55$ & 0.055 & 0.056 & 0.000 \\ 
 
 &  & $\beta_{Z2}$  &   $-0.353$ & 0.152 & 0.159 & 0.944 & $-77.15$ & 0.055 & 0.056 & 0.046 \\ 
 
 & 0.55 & $\beta_{X1}$  &   $-1.095$ & 0.133 & 0.126 & 0.962 & $-84.08$ & 0.027 & 0.027 & 0.000 \\ 
  
 &  & $\beta_{Z1}$  & $-0.866$ & 0.071 & 0.070 & 0.964 & $-51.93$ & 0.045 & 0.044 & 0.015 \\ 
  
 &  & $\beta_{Z2}$  & $-1.897$ & 0.071 & 0.069 & 0.960 & $-40.78$ & 0.045 & 0.047 & 0.337 \\ 
 \hline
\multicolumn{3}{l}{ $Se=0.90, Sp=0.80$} & \multicolumn{4}{c}{Proposed}  & \multicolumn{4}{c}{Naive} \\ \cmidrule(r){4-7} \cmidrule{8-11} $\hat{\delta}_{(1)}$ & CR & $\beta$ & \% Bias & ASE & ESE & CP & \% Bias & ASE & ESE & \multicolumn{1}{c}{CP} \\ 
\hline
0.60 & 0.90 & $\beta_{X1}$  &    $\phantom{-}0.986$ & 0.214 & 0.217 & 0.949 & $-93.42$ & 0.037 & 0.038 & 0.000 \\ 
 &  & $\beta_{Z1}$  &   $\phantom{-}3.516$ & 0.153 & 0.159 & 0.948 & $-87.94$ & 0.046 & 0.048 & 0.000 \\ 
 &  & $\beta_{Z2}$  & $\phantom{-}0.025$ & 0.151 & 0.162 & 0.945 & $-89.97$ & 0.046 & 0.047 & 0.002 \\ 
 & 0.55 & $\beta_{X1}$  &  $-0.488$ & 0.096 & 0.092 & 0.958 & $-78.24$ & 0.034 & 0.034 & 0.000 \\ 
 
 &  & $\beta_{Z1}$  &  $-0.100$ & 0.069 & 0.068 & 0.961 & $-60.97$ & 0.042 & 0.043 & 0.000 \\ 
 
 &  & $\beta_{Z2}$  &  $-0.953$ & 0.068 & 0.067 & 0.957 & $-67.67$ & 0.042 & 0.042 & 0.020 \\

0.30 & 0.90 & $\beta_{X1}$  &     $-0.310$ & 0.301 & 0.303 & 0.951 & $-96.68$ & 0.027 & 0.027 & 0.000 \\

 &  & $\beta_{Z1}$  &    $\phantom{-}2.982$ & 0.161 & 0.167 & 0.952 & $-89.72$ & 0.044 & 0.046 & 0.000 \\

 &  & $\beta_{Z2}$  &   $\phantom{-}0.278$ & 0.160 & 0.170 & 0.941 & $-87.53$ & 0.044 & 0.045 & 0.002 \\ 

 & 0.55 & $\beta_{X1}$  &   $-1.167$ & 0.135 & 0.130 & 0.955 & $-89.05$ & 0.024 & 0.024 & 0.000 \\

 &  & $\beta_{Z1}$  &  $-0.943$ & 0.073 & 0.072 & 0.962 & $-67.08$ & 0.040 & 0.041 & 0.000 \\

 &  & $\beta_{Z2}$  &   $-1.921$ & 0.072 & 0.071 & 0.958 & $-59.89$ & 0.040 & 0.041 & 0.039 \\ 
\hline
\multicolumn{3}{l}{ $Se=1, Sp=1$} &

\multicolumn{4}{c}{Truth}  \\ \cmidrule(r){4-7}   & CR & $\beta$ & \% Bias & ASE & ESE & CP \\
   \hline
 & 0.90 & $\beta_{X1}$ & $\phantom{-}1.652$ & 0.108 & 0.106 & 0.955 \\ 
  
 &  & $\beta_{Z1}$ &  $\phantom{-}2.270$ & 0.108 & 0.110 & 0.949 \\ 
  
&   & $\beta_{Z2}$ & $\phantom{-}0.080$ & 0.108 & 0.112 & 0.949 \\ 
&  0.55 &  $\beta_{X1}$ & $\phantom{-}1.252$ & 0.053 & 0.052 & 0.961 \\

 & &  $\beta_{Z1}$ &   $\phantom{-}1.153$ & 0.053 & 0.052 & 0.961 \\

&   &  $\beta_{Z2}$ &   $\phantom{-}0.223$ & 0.052 & 0.053 & 0.937 \\ 
   \hline
  \end{tabular}
   \begin{tablenotes}
     \item[1] $Se=$ Sensitivity
     \item[2] $Sp=$ Specificity    
     \item[3] $\hat{\delta}_{(1)}=$ estimate of attenuation coefficient
     \item[4] $CR=$ True censoring rate
   \end{tablenotes}
    \end{threeparttable}
    \end{adjustbox}
\end{table}

\begin{table}
    \centering
     \begin{threeparttable}[t]
     \caption{Type I error results for $\beta_{X1}=0$ are given for 1000 simulated data sets for the proposed method. Let $\beta_{X1}=\log(1.5)$, $\beta_{Z1}=\log(0.7)$, and $\beta_{Z2}=\log(1.3)$; $e$ is normally distributed with mean zero.}
    \label{T1errtab}
\begin{tabular}{llllc}
\hline
$Se\tnote{1}$ & $Sp\tnote{2}$ & $\hat{\delta}_{(1)}$\tnote{3} & CR\tnote{4} & \multicolumn{1}{c}{Type I Error} \\ 
\hline
 0.80 & 0.90 & 0.30 & 0.55  & $0.048$ \\
 &  &  & 0.90  & $0.042$ \\
 &  & 0.60 & 0.55  & $0.058$ \\
 &  &  & 0.90  & $0.042$ \\
0.90 & 0.80 & 0.30 & 0.55  & $0.043$ \\
 &  &  & 0.90  & $0.039$ \\
 &  & 0.60 & 0.55  & $0.049$ \\
 &  &  & 0.90  & $0.044$ \\
\hline 
\end{tabular}
 \begin{tablenotes}
     \item[1] $Se=$ Sensitivity
     \item[2] $Sp=$ Specificity    
     \item[3] $\hat{\delta}_{(1)}=$ estimate of attenuation coefficient
     \item[4] $CR=$ True censoring rate
   \end{tablenotes}
    \end{threeparttable}
\end{table}

% latex table generated in R 3.5.2 by xtable 1.8-3 package
% Thu Sep 12 11:45:50 2019
\begin{table}
\centering
 \begin{threeparttable}[t]
\caption {Hazard Ratio (HR) and 95\% confidence interval (CI) estimates of incident diabetes for a 20\% increase in consumption of energy (kcal/d), protein (g/d), and protein density (\% energy from protein/d) based on the naive method ignoring error in the outcome and covariate, the regression calibration method that corrects for covariate error only, and the proposed method. Here, sensitivity = 0.61, specificity = 0.995, and negative predictive value = 0.96.} \label{tableHR} 

\centering
\begin{tabular}{llll}
  \hline
  \multicolumn{2}{c}{} & \multicolumn{2}{c}{HR (95\% CI)} \\  \multicolumn{1}{l}{Model\tnote{1}} & \multicolumn{1}{l}{Method} & \multicolumn{1}{c}{Adjusted for BMI\tnote{2}}  & \multicolumn{1}{c}{Not Adjusted for BMI} \\  
  \hline
 Energy (kcal/d) &  Naive  & 1.002 (0.986, 1.018) & 1.024 (1.008, 1.040) \\ 
  & Regression Calibration & 1.041 (0.758, 1.429) & 1.421 (1.043, 1.938) \\ 
  & Proposed & 0.822 (0.512, 1.318) & 1.189 (0.836, 1.692) \\   \hline

   Protein (g/d) &

  Naive & 1.024 (1.010, 1.039) & 1.051 (1.035, 1.066) \\ 
  & Regression Calibration & 1.121 (1.036, 1.213) & 1.231 (1.130, 1.342) \\ 
 & Proposed & 1.077 (0.978, 1.186) & 1.241 (1.114, 1.384) \\   \hline

  Protein Density & Naive  & 1.100 (1.064, 1.137) & 1.128 (1.091, 1.167) \\ 
 &  Regression Calibration  & 1.243 (1.125, 1.374) & 1.325 (1.181, 1.486) \\ 
 & Proposed & 1.266 (1.115, 1.436) & 1.327 (1.183, 1.490) \\ 
   \hline
\end{tabular}
 \begin{tablenotes}
     \item[1] Each model is adjusted for potential confounders and is stratified on age (10-year categories) and Dietary Modification trial (DM) or Observational Study (OS) cohort membership.
     \item[2] BMI = Body Mass Index $(kg/m^2)$
   \end{tablenotes}
    \end{threeparttable}%
\end{table}

\newpage

\begin{center}
{\Large \textbf{Supplementary Materials for ``An Approximate Quasi-Likelihood Approach for Error-Prone Failure Time Outcomes and Exposures"} }
\end{center}

\begin{center}
\normalsize{Lillian A. Boe$^{1,*}$, 
Lesley F. Tinker$^{2}$, and 
Pamela A. Shaw$^{1}$ \\
$^{1}$Department of Biostatistics, Epidemiology, and Informatics, \\ University of Pennsylvania Perelman School of Medicine,  Philadelphia, PA 19104\\
$^{2}$WHI Clinical Coordinating Center,\\ Fred Hutchinson Cancer Research Center, Seattle, WA 98109 \\
\textit{*email}: boel@pennmedicine.upenn.edu}

\end{center}

\section*{Contents}
\begin{itemize}
    \item  \hyperref[Rcode]{\ref{Rcode}: R Code with illustrative example} 
    \item \hyperref[deriv3]{S2: Derivation of equation (2)}
\item \hyperref[theory]{S3: Regularity conditions}
\item \hyperref[dataset]{S4. Supplemental methods and discussion for Women's Health Initiative data} example
\item \hyperref[fig:my_label]{Supplemental figures}
\item \hyperref[table1simsappendix]{Supplemental tables}
\end{itemize}

\setcounter{section}{0}
\renewcommand{\thesection}{S\arabic{section}}

\section{R Code with illustrative example}\label{Rcode}

In this section, we provide an illustrative example showing how we can use R code to apply the proposed method. For this example, we will use a simulated data set with one error-prone covariate ($X_1^{*}$) and 4 precisely-recorded covariates ($Z_1$, $Z_2$, $Z_3$, $Z_4$). Additionally, we have periodic follow-up from 4 visits. The sensitivity, specificity, and negative predictive value of the error-prone outcome are assumed to be $0.60$, $0.98$, and $0.95$, respectively. This data set can be found on GitHub at \href{https://github.com/lboe23/Outcome-Error-RC}{https://github.com/lboe23/Outcome-Error-RC} under the file name Simulated\_Data\_Example\_5Cov\_Long.csv.

Before we begin our analysis, we need to load the Rcpp functions that we need to compute the likelihood and the function for the variance calculation. We use the following Rcpp functions, which were developed by \citet{gu2015semiparametric} and can be found in the icensmis package on Cran:  ``loglikC," ``gradlikC," ``dmat" and ``getrids." The variance calculation function can be found on the GitHub site listed above under the file name ``Variance\_functions.R." 

\begin{figure}[H]
    \centering
    \includegraphics[width=1\textwidth]{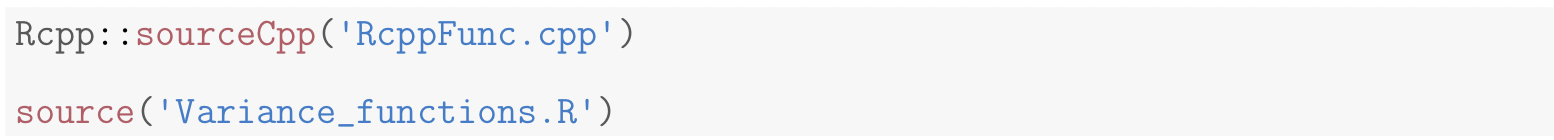}
\end{figure}

As suggested by the name of our data set, the input data is in long form, where each row represents one time point and each subject has multiple rows. Below we read our data into R and then present the first 6 rows of the data.

\begin{figure}[H]
    \centering
    \includegraphics[width=1\textwidth]{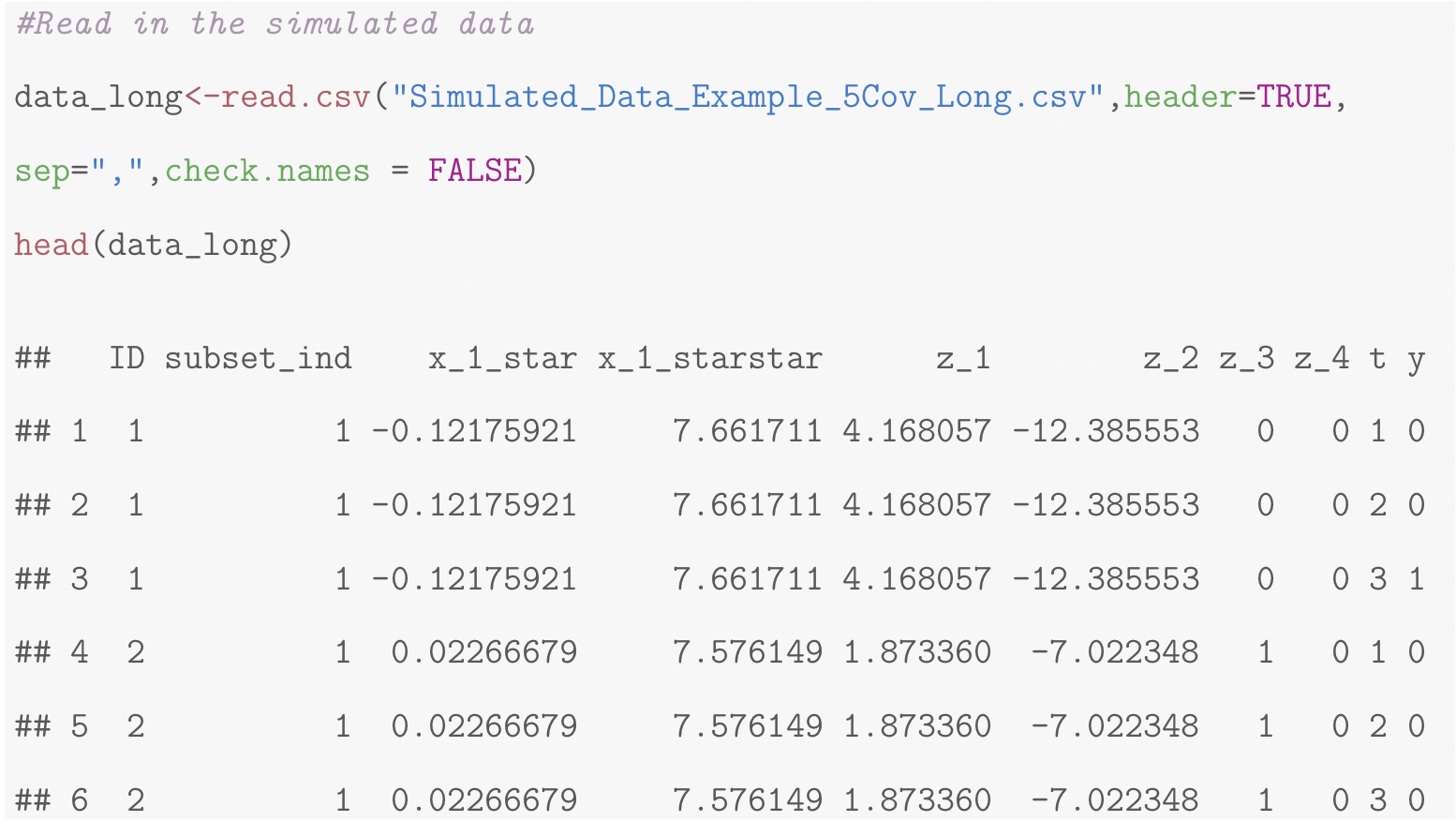}
\end{figure}

Our input dataset consists of the following variables:

\begin{itemize}
    \item $ID$: a unique ID for each subject in the data set.
     \item $subset\_ind$: an indicator variable representing membership in the calibration subset, which takes the value 1 if the subject is a member of the calibration subset and 0 if they are not. In this data example, $n_C=500$ of the $N=10,000$ total study subjects are in the calibration subset. 
    \item $x\_1\_star$: the error-prone covariate of interest, prone to both systematic and random error (e.g. self-reported measure of dietary energy).
     \item $x\_1\_starstar$: the covariate of interest subject to classical measurement error (e.g. biomarker of dietary energy), which is only available for members of the validation subset (those with $subset\_ind=1$). 
    \item $z\_1, z\_2$: precisely measured continuous covariates that we wish to include in our calibration and outcome models.
    \item $z\_3, z\_4$: precisely measured binary covariates that we wish to include in our calibration and outcome models.
    \item $y$: the error-prone, binary result where 1 indicates a positive test and 0 indicates a negative test.
    \item $t$: the visit time corresponding to each error-prone test result, y.
\end{itemize}

To apply the proposed method, we will begin by fitting the calibration equations. First, we create a new dataset that only has one row per subject and only includes the members of the calibration subset.

\begin{figure}[H]
    \centering
    \includegraphics[width=1\textwidth]{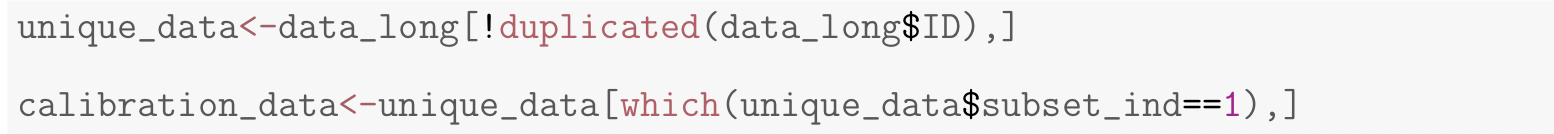}
\end{figure}

Next, we will fit the calibration model by regressing the covariate measure with classical measurement error, $X_1^{**}$ on the covariate with prone to more extreme error, $X^*$, and other covariates, $Z_1$, $Z_2$, $Z_3$, and $Z_4$. We note that the model below corresponds to equation \ref{xdoublestar} of in main manuscript: $X^{**}_i=\delta_{(0)}+\delta_{(1)}X_i^{*}+\delta_{(2)}Z_i+V_i$.

\begin{figure}[H]
    \centering
    \includegraphics[width=1\textwidth]{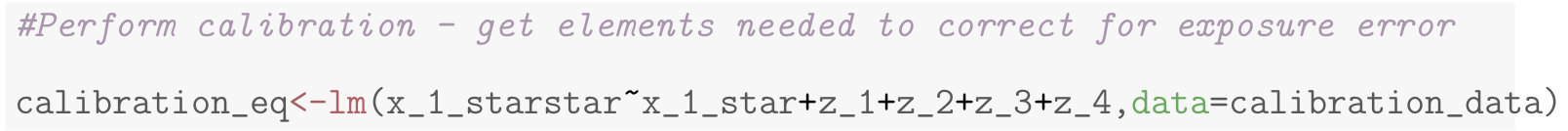}
\end{figure}

We will now save the summary data from the calibration equation and use this to create our multivariate correction factor from equation \ref{multicorrection} of the main manuscript, which recall has the following form: 

\begin{equation*}\label{multicorrectioncode}
\hat{\Delta}=\begin{bmatrix}
\hat{\delta}_{(1)p\times p} & \hat{\delta}_{(2)p\times q} \\
0_{q\times p} & I_{q\times q} 
\end{bmatrix}_.
\end{equation*}

\begin{figure}[H]
    \centering
    \includegraphics[width=1\textwidth]{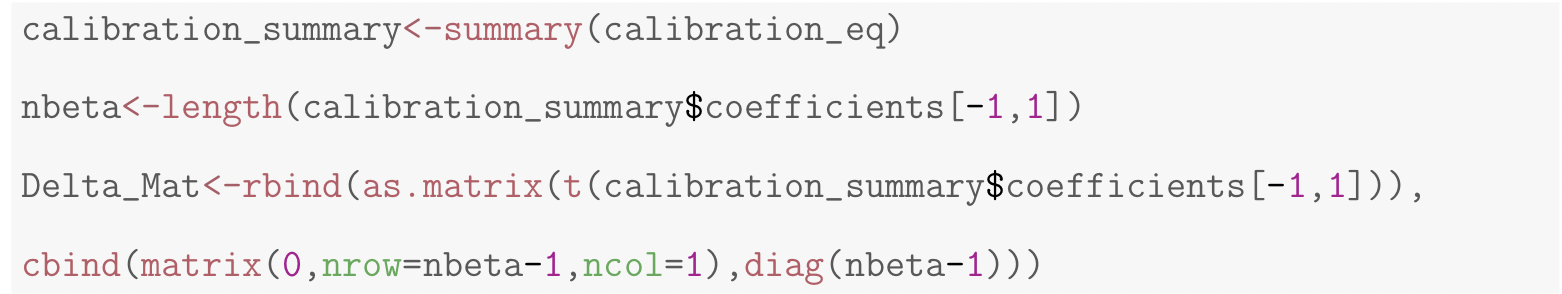}
\end{figure}

Next, we want to save the elements of the variance-covariance matrix from the calibration equation, as this will be used later in the computation of the variance-covariance matrix $\Sigma$ for $\hat{\beta}$. Note that we do not need the elements of the variance-covariance matrix that correspond to the intercept term for this approach.

\begin{figure}[H]
    \centering
    \includegraphics[width=1\textwidth]{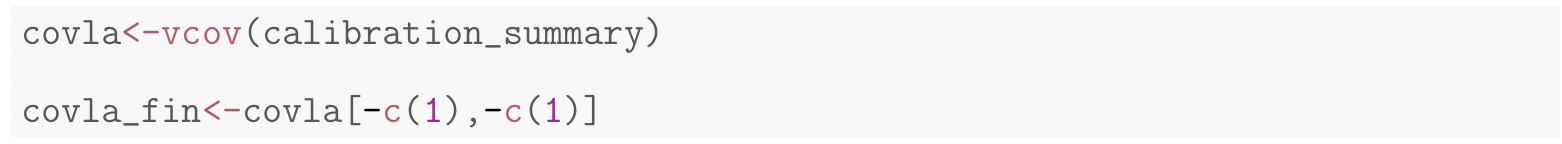}
\end{figure}

We will now begin the process of fitting our outcome models. As we did for the WHI data example in the text, we will consider 3 approaches: (1) the naive method ignoring error in the outcome and covariate, (2) the regression calibration method that corrects for error in the covariate only, and (3) the proposed method. First, let's assign sensitivity, specificty, and negative predictive value. 

\begin{figure}[H]
    \centering
    \includegraphics[width=1\textwidth]{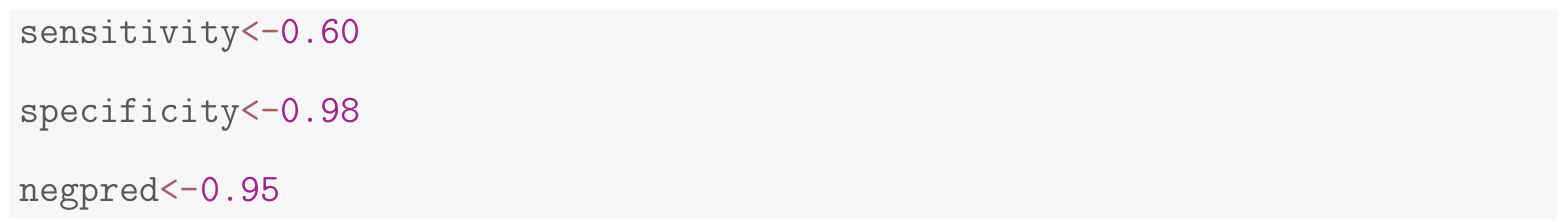}
\end{figure}

We will now fit our first outcome model, which corresponds to the naive approach. To estimate regression coefficients for the naive grouped continuous time Cox proportional hazards model, we will fit a generalized linear model with a binomial outcome and assume a complementary log-log link. 

\begin{figure}[H]
    \centering
    \includegraphics[width=1\textwidth]{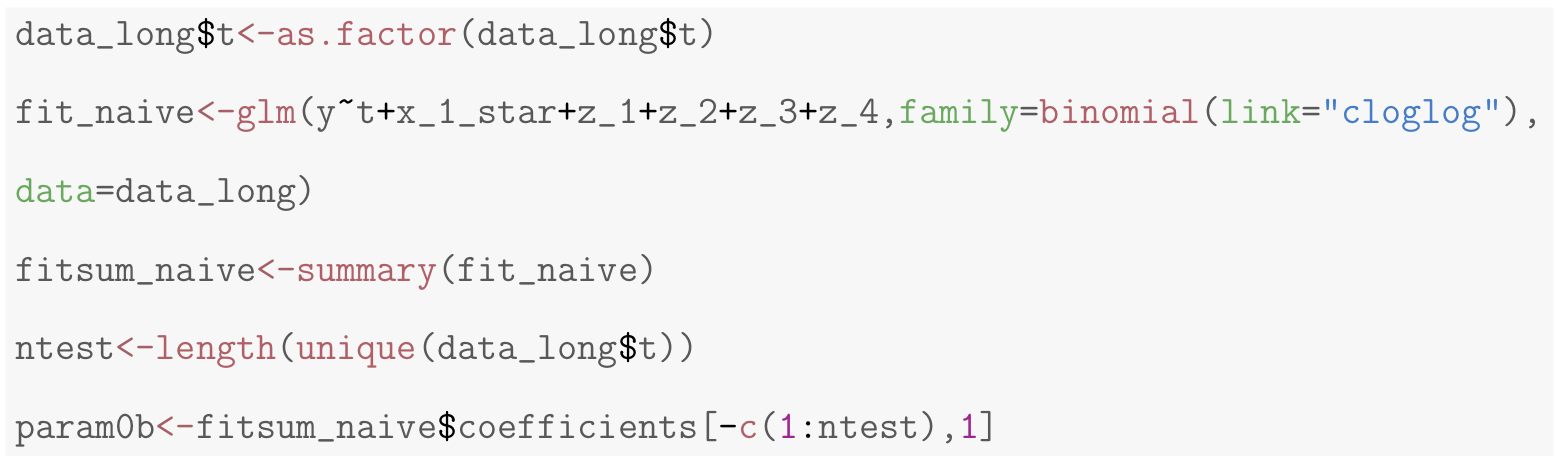}
\end{figure}

Now we will get our data in the format required to use the proposed method. First, we will define the formula that we want to use for our outcome model. 

\begin{figure}[H]
    \centering
    \includegraphics[width=1\textwidth]{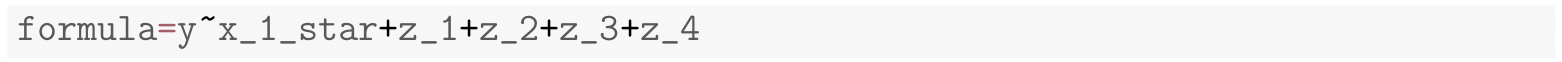}
\end{figure}

Now, let's make sure our data is ordered properly before we begin calculating sum of the likelihood components. 

\begin{figure}[H]
    \centering
    \includegraphics[width=1\textwidth]{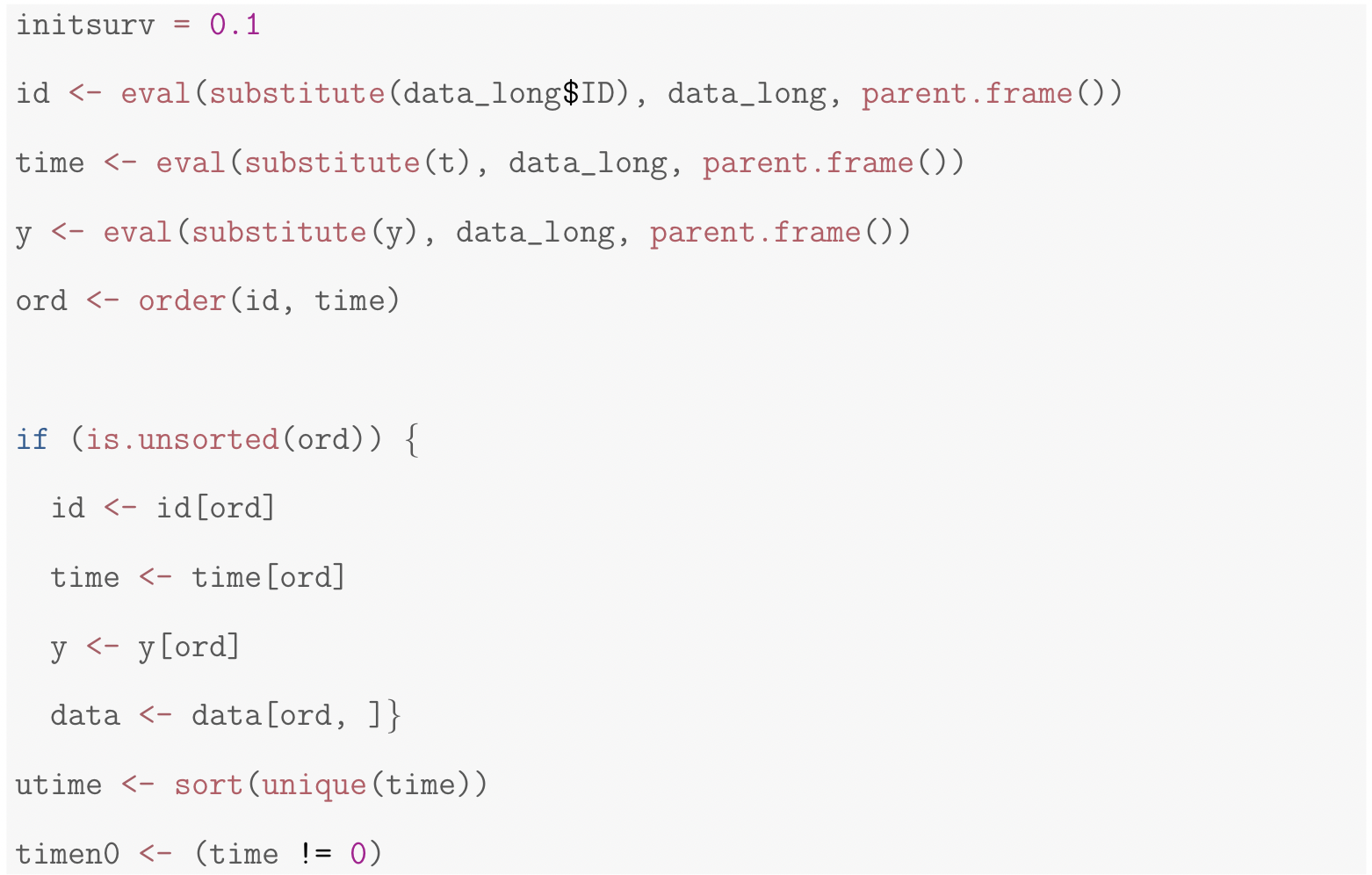}
\end{figure}

Now that our data is in an appropriate form, we can calculate the $D$ matrix, defined in section \ref{ref_section1} of the main manuscript. Additionally, we calculate $J$ and the number of rows in $D$.

\begin{figure}[H]
    \centering
    \includegraphics[width=1\textwidth]{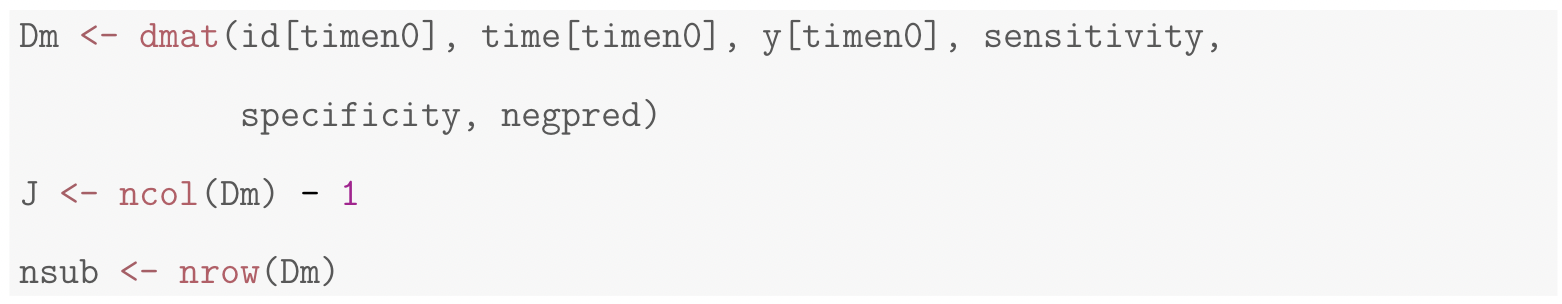}
\end{figure}

As we get ready to maximize our log-likelihood, we want to think of starting values for our survival parameters. To avoid maximization problems due to the ordered constraint of the survival parameters $1=S_1>S_2>...>S_{J+1}>0$, we re-parameterize these terms for optimization. The re-parameterization that we use is a log-log transformation of
survival function for $S_2$, and a change in log-log of the survival function for all other parameters. We consider initial values of 0.1 for our survival parameters, then transform these based on this re-parameterization. Additionally, we define a lower bound of $-\infty$ for the first re-parameterized survival function and 0 for the subsequent $J-1$ terms. 

\begin{figure}[H]
    \centering
    \includegraphics[width=1\textwidth]{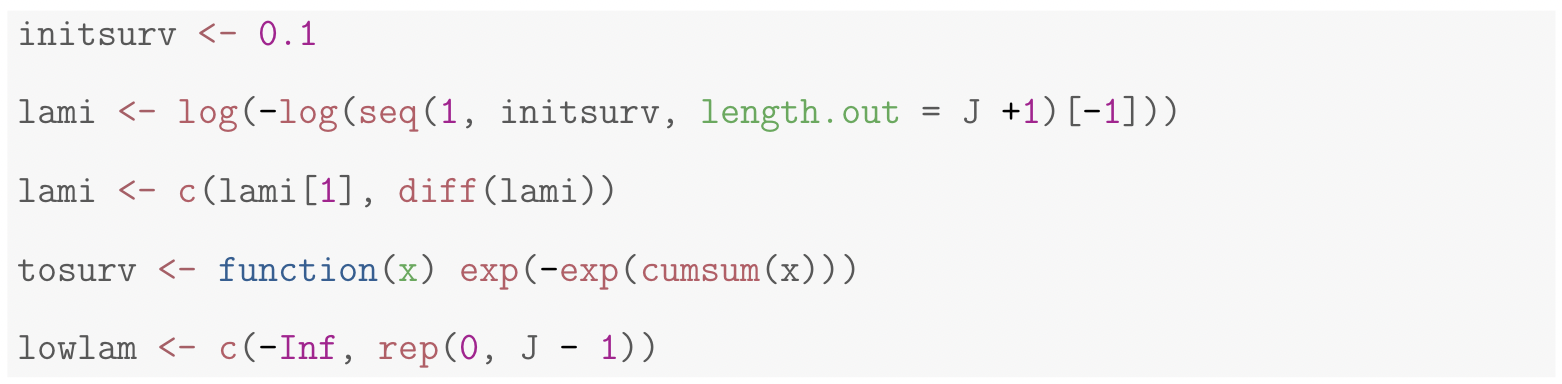}
\end{figure}

Next, we want to create a matrix version of our covariate data which will be used in the maximization of the log-likelihood.

\begin{figure}[H]
    \centering
    \includegraphics[width=1\textwidth]{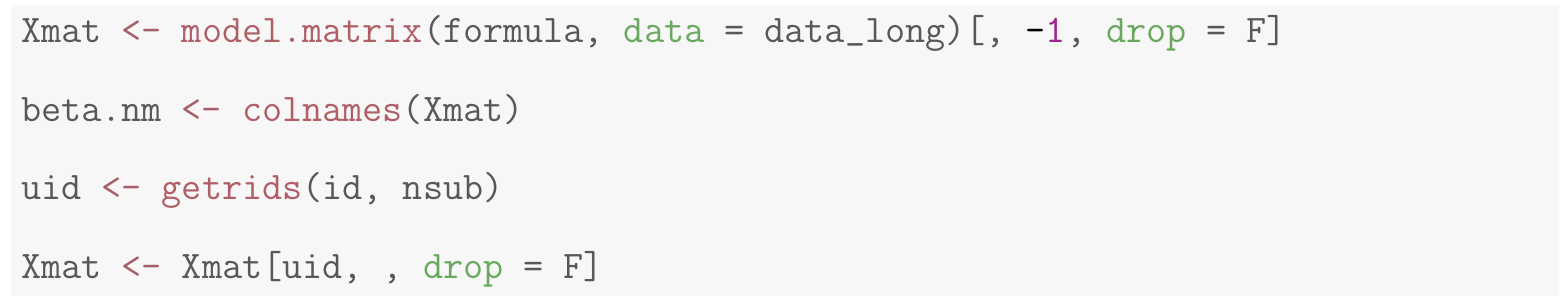}
\end{figure}

We will now maximize our log-likelihood function that corrects for outcome error only using the ``L-BFGS-B" method in the optim function. We will give the lower bound $lowlam$ defined above for our survival function parameter estimates and a lower bound of $\infty$ for our regression coefficient estimates. We will use the $lami$ values defined above as our initial values for our baseline survival functions. We will use the estimated regression parameters from the naive method as our starting values for $\beta_{X1}$, $\beta_{Z1}$, $\beta_{Z2}$, $\beta_{Z3}$, and $\beta_{Z4}$ in the proposed method. 

\begin{figure}[H]
    \centering
    \includegraphics[width=1\textwidth]{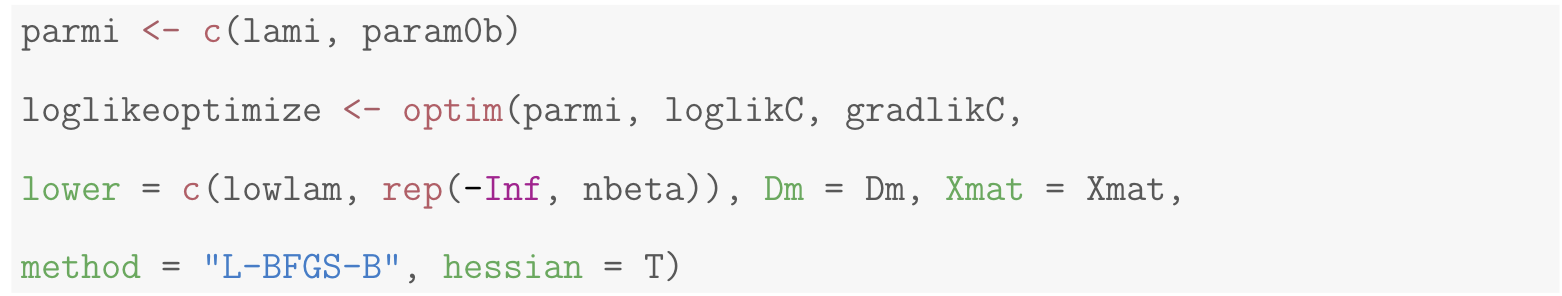}
\end{figure}

We can now  invert the Hessian matrix to calculate $\hat{\Sigma}_{\beta^*}$.

\begin{figure}[H]
    \centering
    \includegraphics[width=1\textwidth]{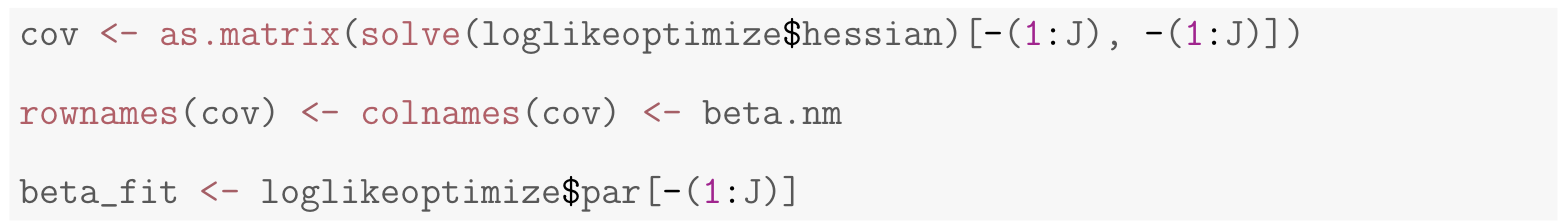}
\end{figure}

It is finally time to apply the proposed method. Below, we calculate our corrected vector of estimated regression coefficients of interest, using equation \ref{posthoccorr} from the main manuscript: $\hat{\beta}=\hat{\beta}^*\hat{\Delta}^{-1}$. Recall that we computed $\hat{\Delta}$ above using regression calibration. 

\begin{figure}[H]
    \centering
    \includegraphics[width=1\textwidth]{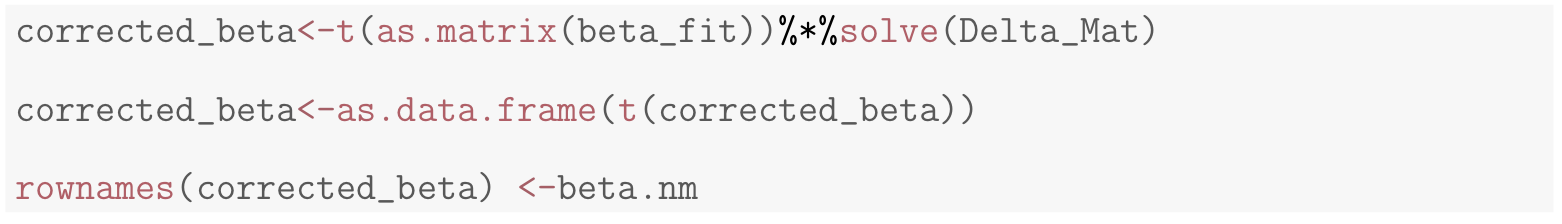}
\end{figure}

Lastly, we compute the variance for the proposed approach. To do this, we use the function ``Proposed\_Var" from the Variance\_functions.R file that we imported above. This code for the variance calculation can accommodate 1 error-prone covariate and up to 19 precisely-measured covariates, for a total of 20 covariates in the calibration and outcome models. The input values for this function, in order, are the following: (1) $\hat{\Sigma}_{\beta^*}$, the variance-covariance matrix from the method that corrects for outcome error only; (2) the variance-covariance matrix from the calibration model; (3) the estimated multivariate correction factor from regression calibration, $\hat{\Delta}$; and (4) the estimated regression parameters obtained by fitting the model that corrects for outcome error only.

\begin{figure}[H]
    \centering
    \includegraphics[width=1\textwidth]{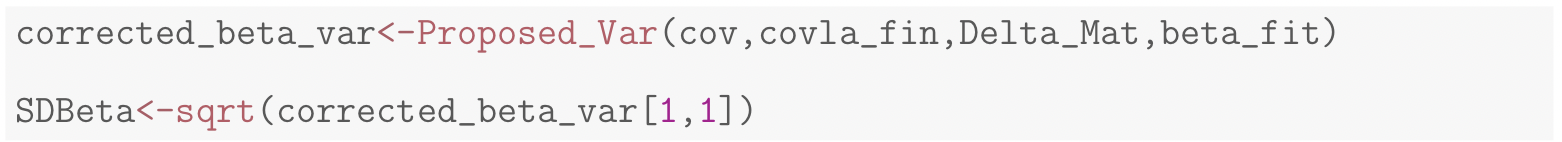}
\end{figure}

Now, to complete our results table, we will use regression calibration to obtain the results for the method that corrects for covariate error only:

\begin{figure}[H]
    \centering
    \includegraphics[width=1\textwidth]{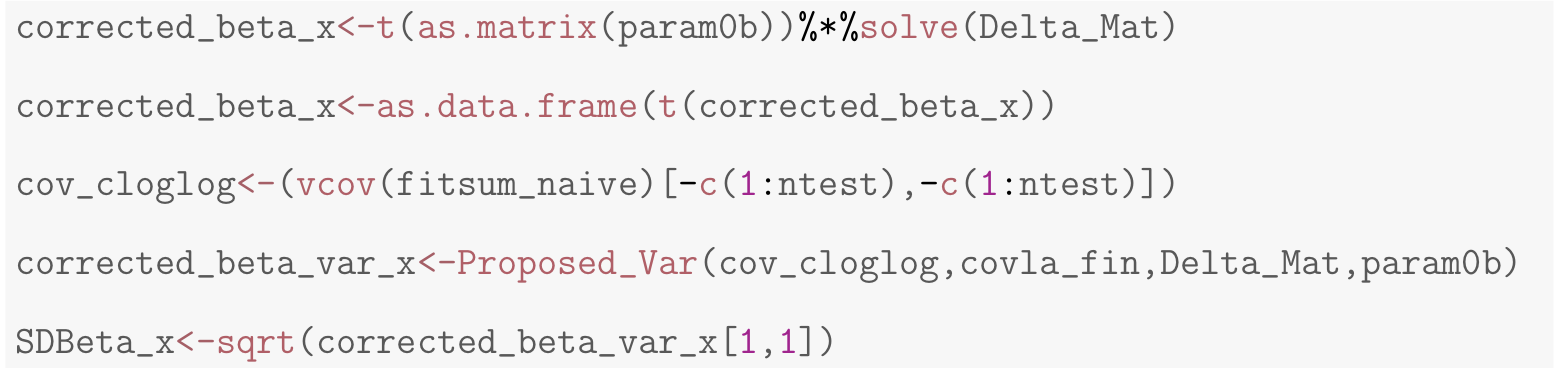}
\end{figure}

The last step is to exponentiate our regression parameters and corresponding confidence interval bounds and put them into a table so that we can present the results for all three methods simultaneously. The final results are presented below:

\begin{figure}[H]
    \centering
    \includegraphics[width=1\textwidth]{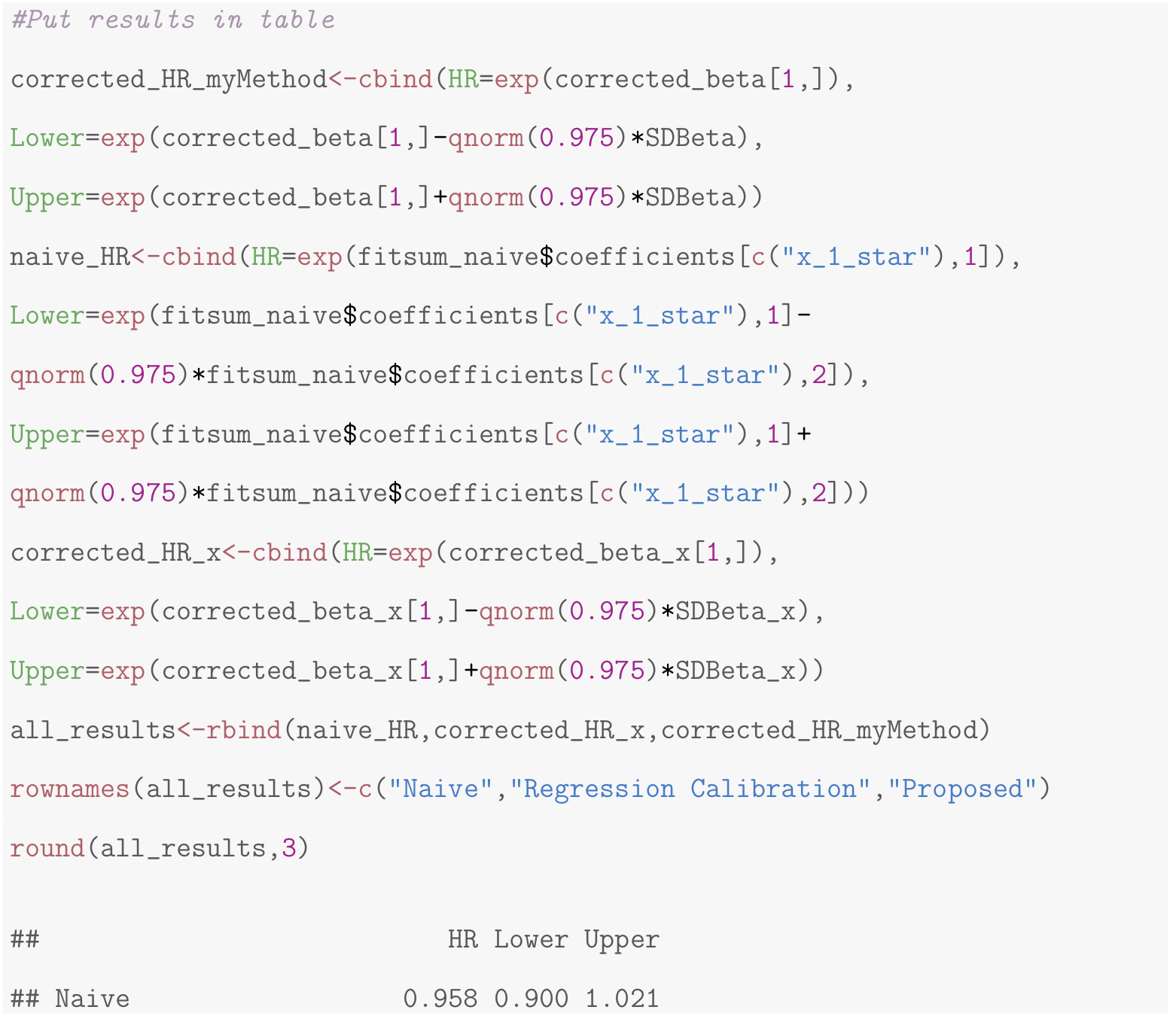}
\end{figure}
\begin{figure}[H]
    \centering
    \includegraphics[width=1\textwidth]{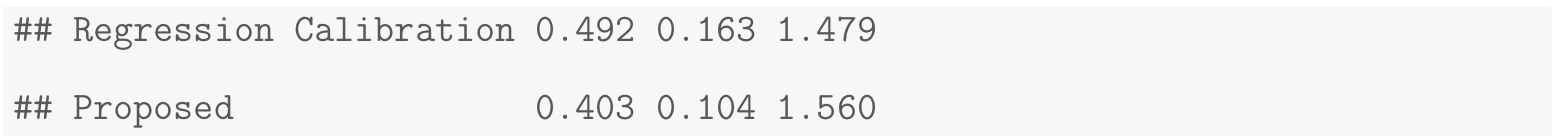}
\end{figure}

\section{Derivation of equation \ref{jointprob}}\label{deriv3}

In this section, we follow the notation and logic of \citet{balasubramanian2003estimation} to show how we derive the likelihood contribution for subject $i$ in equation \ref{jointprob} from equation \ref{firstequation}. The steps are as follows:
\begin{eqnarray*}
    f(\mathbf{Y_i},\mathbf{t_i},n_i)&=&\sum_{j=1}^{J+1}\Pr(\tau_{j-1}<T_i\leq\tau_j) \Pr(\mathbf{Y_i},\mathbf{t_i},n_i| T_i), \\
    &=&\sum_{j=1}^{J+1}\theta_j \prod_{l=1}^{n_i}\Pr(t_{il},Y_{il}|t_{i1},t_{i2},\ldots,t_{il-1},Y_{i1},Y_{i2},\ldots,Y_{il-1},T_i)  \\ && \times \Pr(n_i|t_{i1},t_{i2},\ldots,t_{in_i},Y_{i1},Y_{i2},\ldots,Y_{in_i},T_i) \\
    &=&\sum_{j=1}^{J+1}\theta_j \prod_{l=1}^{n_i}\Pr(t_{il}|t_{i1},t_{i2},\ldots,t_{il-1},Y_{i1},Y_{i2},\ldots,Y_{il-1},T_i)  \\ && \times
    \prod_{l=1}^{n_i}\Pr(Y_{il}|t_{i1},t_{i2},\ldots,t_{il-1},t_{il},Y_{i1},Y_{i2},\ldots,Y_{il-1},T_i)  \\ && \times \Pr(n_i|t_{i1},t_{i2},\ldots,t_{in_i},Y_{i1},Y_{i2},\ldots,Y_{in_i},T_i)
\end{eqnarray*}
\noindent where $\theta_j=\Pr(\tau_{j-1}<T_i\leq\tau_j)$. Now, by the assumption that $\Pr(\mathbf{Y_i}|T_i,\mathbf{t_i})=\prod_{l=1}^{n_i}\Pr(Y_{il}|T_i,t_{il})$:

\begin{eqnarray*}
    f(\mathbf{Y_i},\mathbf{t_i},n_i)
    &=&\sum_{j=1}^{J+1}\theta_j \prod_{l=1}^{n_i}\Pr(t_{il}|t_{i1},t_{i2},\ldots,t_{il-1},Y_{i1},Y_{i2},\ldots,Y_{il-1},T_i)  \times
    \prod_{l=1}^{n_i}\Pr(Y_{il}|T_i,t_{il}) \\ && \times \Pr(n_i|t_{i1},t_{i2},\ldots,t_{in_i},Y_{i1},Y_{i2},\ldots,Y_{in_i},T_i)
\end{eqnarray*}

Finally, following Balasubramanian and Lagakos, \cite{balasubramanian2003estimation} for the case of a prespecified visit schedule, we have the following:

\begin{eqnarray*}
   &&  \Pr(t_{il}|t_{i1},t_{i2},\ldots,t_{il-1},Y_{i1},Y_{i2},\ldots,Y_{il-1},T_i)  \\ && = \Pr(n_i|t_{i1},t_{i2},\ldots,t_{in_i},Y_{i1},Y_{i2},\ldots,Y_{in_i},T_i) = 1
\end{eqnarray*}

\noindent Note, these two probabilities would also drop out of the likelihood if they did not depend on the parameters of interest ($\beta$). Now, we arrive at equation \ref{jointprob} for the likelihood contribution for the $i$th subject: 

\begin{eqnarray*}
    f(\mathbf{Y_i},\mathbf{t_i},n_i)
    &=&\sum_{j=1}^{J+1}\theta_j 
    \prod_{l=1}^{n_i}\Pr(Y_{il}|\tau_{j-1}<T_i\leq \tau_j,t_l) =\sum_{j=1}^{J+1}\theta_j 
    C_{ij}
\end{eqnarray*}

\noindent where $C_{ij}=\prod_{l=1}^{n_i}\Pr(Y_{il}|\tau_{j-1}<T_i\leq \tau_j,t_l).$

\section{Regularity conditions}\label{theory}

In this section, we outline sufficient regularity conditions for the proposed estimator, namely asymptotic normality and $\sqrt{N}$-convergence. Recall that we have an approximate estimator that has empirically been observed to have good properties, i.e. have minimal bias and close to nominal coverage, when the event of interest is rare and the true parameter value is of moderate size. 

First assume that we have discrete observation times for the failure time that satisfy the following: $0=\tau_0<\tau_1<\tau_2<...<\tau_J <\tau_{J+1} = \infty$. Further, define the elements of $t_i$, the vector of visit times for subject $i$, to be a subset of $\{\tau_0,\tau_1,\tau_2,...,\tau_J\}$. Recall that we define $S_j=\Pr(T>\tau_{j-1})$ for $j=1,...,J+1$; and $\tau_0 = 0$, and  and require that $1=S_1>S_2>...>S_{J+1}>0$. The previous two conditions ensure that $0 < \theta_j < 1$ for $j=1,...,J$, where $\theta_j=\Pr(\tau_{j-1}<T\leq\tau_j)$. Now, assume the following: (1) $\{X_i, X^{*}_i,Z_i,T_i,Y_i,t_i\}$, $i=1,...,N$ are independent and identically distributed, where $N$ is the number of subjects in the main study data; and (2) $\frac{n_C}{N} \rightarrow p\in (0,1)$, where $n_C$ is the number of subjects in the calibration subset.
 
Assume that $\{X_i,Z_i,Y_i,t_i\}$, for $i=1,...,N$ follows the density $ f(\mathbf{X_i, Z_i,Y_i,t_i};\psi^0)$ with the corresponding log-likelihood function $l(\psi)=l(X_i,Z_i,Y_i,t_i;\psi)$; where $\mathbf{\psi=[\beta,S]}$,  $\beta=(\beta_X,\beta_Z)^T$, $\mathbf{S} = (S_1,S_2,...,S_{J+1})^T$, and $l(\psi)$ is as defined in equation \ref{likelihood} of the main text, i.e.

\begin{equation}
    l(\psi)=l(\mathbf{S},\beta)=\sum_{i=1}^{N}\log\left(\sum_{j=1}^{J+1}D_{ij}S_j^{\exp(x_i^T\beta_X+z_i^T\beta_Z)}\right).
\end{equation}
  
\noindent Here, $\psi^0$ is the vector of regression parameters of interest for the likelihood with the unobserved true data for $X$. Assume the log-likelihood is twice continuously differentiable and define $l^\star(\psi)= l(X^*_i,Z_i,Y_i,t_i;\psi)$.  Let the partially naive score function be denoted $U_N^*({\psi})=(1/N)\partial l^{\star}(\psi)/\partial \psi$, and let $\hat{\psi}_N^*$ to be the solution to the score equations, $U_N^*({\psi})=0$.  Define $\psi^*$ to be the vector of parameters that solves $E\left[\frac{\partial l(X^*,Z,Y,t;\psi)}{\partial \psi}\right]=E[U^*(\psi)]=0$.  $\psi^*$ will not generally be equal to $\psi^0$, since the partially naive likelihood does not adjust for the covariate error in $X^*$.  Because $\hat{\psi}_N^*$ is a maximum likelihood estimator (MLE), we can rely on standard regularity conditions to see that with probability going to one as $N \rightarrow \infty$, there exists a unique solution to the likelihood equations, $\hat{\psi}_N^*$, that is consistent for $\psi^*$ \citep{foutz1977unique} and asymptotically normal.\citep{boos2013essential} Under these regularity conditions, one has
\begin{equation}
    \sqrt{N}(\hat{\psi}_N^*-\psi^*) \xrightarrow{d} \mathcal{N}(0, I(\psi^*)^{-1}),
\end{equation}

\noindent where $I(\psi^*)^{-1}$ is the Fisher information matrix.

Recall that the proposed estimator $\hat{\beta}$ is defined as $\hat{\beta}^*\hat{\Delta}^{-1}$, where $\hat{\beta}^*$ is the first $p+q$ elements of the vector $\hat{\psi}_N^*$. Since $\hat{\Delta}$ is a linear regression estimator, we can also appeal to standard MLE theory to establish its consistency for the true parameter $\Delta^0$ and asymptotic normality.  Finally, we need only satisfy the necessary regularity conditions for the multivariate delta method to establish consistency and asymptotic normality of our proposed estimator. In addition to the established asymptotic normality of $(\hat{\beta},\hat{\Delta})$, for $g(\beta,\Delta)=\beta\Delta^{-1}$, we need only that its matrix of partial order derivatives be continuous in a neighborhood of $(\beta^*,\Delta^0)$.\citep{casella2002statistical} Further assume the independence of $\hat{\beta}$ and $\hat{\Delta}$, which holds if the  number of subjects in the calibration subset, $n_C$, is a small percentage of the main study sample size, $N$. Then, we have:
\begin{equation}
   \sqrt{N}\left(\hat{\beta}^*\hat{\Delta}^{-1} -\beta^*(\Delta^0)^{-1}\right) \xrightarrow{d}  \mathcal{N}(0, \Sigma),
\end{equation}

\noindent where the $(j_1,j_2)^{th}$ element of $\Sigma$ is defined as $\Sigma_{\beta}(j_1,j_2)\cong\left(A'\Sigma_{\beta^*}A\right)_{j_1,j_2}+\beta^*\Sigma_{A,j_1,j_2}\beta^{*\prime}$, with $\Sigma_{\beta^*}$ the asymptotic variance of $\hat{\beta}$, and $A=\Delta^{-1}$ and $\Sigma_{A,j_1,j_2}$ defined similarly as in the main text. 

The numerical performance of our proposed estimator has been studied extensively and  shown to perform well empirically, as described in the main manuscript. Under these standard regularity conditions, we have illustrated the asymptotic normality of our estimator in the context of an error-prone time-to-event outcome and covariate.

\section{Supplemental methods and discussion for Women's Health Initiative data example}\label{dataset}

We adopted exclusion criteria in order to obtain a final analytic data set for our analyses that approximated that used by Tinker et al. \cite{tinker2011biomarker} Applying these exclusion criteria resulted in approximately the same cohort. We excluded anyone who reported diabetes at baseline or during the first year of follow-up for the comparison arm of the WHI Dietary Modification trial (DM-C) participants ($n=724$) or the first three years of follow-up for the WHI Observational Study (OS) participants ($n=4109$). We attempted to align characteristics of participants in the DM-C trial with those of participants in the OS by excluding the following participants in the OS: those who had breast, colorectal, or other cancer within 10 years prior to enrollment ($n=8677)$, stroke or myocardial infarction within 6 months prior to enrollment ($n=155$), body mass index (BMI) $<18$ ($n=678$), hypertension (systolic blood pressure $>200$ or diastolic blood pressure $>105$)($n=244$), reported daily energy intake of $<600$ kcal or $>5000$ kcal ($n=3571$), $\geq 10$ meals prepared away from home each week ($n=3598$), a special low-fiber diet ($n=568$), a special malabsorption-related diet ($n=514$), inadvertent weight loss of $>15$ pounds within 6 months of enrollment ($n=594$), and reported diabetes diagnosis before age 21 at enrollment ($n=95$). Applying these exclusion criteria and selecting only the participants with no missing data in the calibration and outcome model variables, we arrived at our analytic cohort with 65,358 members. Of these 65,358 participants, 12,121 (18.5\%) were from the DM-C and 53,237 (81.5\%) were from the OS.

We note that our HRs for the case of correcting for covariate error only were substantially different than those originally reported by Tinker et al. \cite{tinker2011biomarker} For example, \citet{tinker2011biomarker} reports that a HR (95\% CI) of 2.41 (2.06, 2.82) was associated with a 20\% increase in energy intake when BMI was omitted from the outcome model, compared to our 1.421 (1.043, 1.938). There were several differences between these analyses that may have led to this, including slightly different data sets. We reanalyzed our data using a continuous Cox model and found results that were very consistent results with our discrete analysis, so the discrete approach did not explain this difference. First, we investigated the potential discrepancies in results that might arise from the choice of strata. In our original analysis, we stratified our models on age in 10-year categories and DM-C or OS cohort membership, which resulted in 6 strata. We used a continuous Cox model to assess how our results changed when we expanded our strata to (1) age in 5-year categories and DM-C or OS cohort membership (12 strata) or (2) age in 5-year categories, DM-C or OS cohort membership, and hormone therapy trial arm (active estrogen, estrogen placebo, active estrogen plus progestin, estrogen plus progestin placebo, and not randomized) for participants in the DM-C who were also on the hormone trials (36 strata). Table \ref{stratsens} compares our original results using the discrete proportional hazards model and correcting for covariate error to the results using the continuous time Cox proportional hazards model and allowing for either the 6, 12, or 36 strata described above. When we used a Cox model and applied the post-hoc regression calibration approach to correct for covariate error, we obtained the following HR (95\% CI) for a 20\% increase in energy intake when the model did not adjust for BMI: 6 strata, 1.333 (0.993, 1.790); 12 strata, 1.334 (0.994, 1.791); 36 strata, 1.328 (0.990, 1.780). Note that these results are fairly consistent with those obtained for the discrete model correcting for covariate error only (HR 1.421; 95\% CI 1.043, 1.938). Furthermore, we see that our results were not sensitive to the choice of strata.

One important difference between analyses is that we aligned the covariates between the outcome and calibration models, but \citet{tinker2011biomarker} did not.  This alignment is necessary for our approach and in general is recommended for regression calibration in order to avoid potential sources of bias.\citep{kipnis2009modeling} We used a continuous model and a traditional regression calibration approach (non-post-hoc) to show how the results that correct for covariate error only might differ based on the following: BMI is in (1) both the calibration and outcome model, (2) neither model, or (3) the calibration model only. The latter case is an example of not aligning the calibration and outcome model and is not possible for our post-hoc approach used for correcting covariate error. Results comparing these different alignment strategies are presented in table \ref{supptablestrat}. This table presents HR estimates and 95\% confidence intervals for the discrete analysis with the post-hoc correction for covariate error, the continuous Cox model analysis with the post-hoc correction for covariate error, and the continuous Cox model analysis with the non-post-hoc traditional regression calibration correction for covariate error. For analyses that include BMI in both the calibration and outcome models, we obtain similar results for all three approaches for energy, protein, and protein density, indicating that the choice of a discrete analysis or a post-hoc correction does not substantially change our answer. The same is true for analyses that include BMI in neither the calibration nor the outcome model. As we saw in the main manuscript, adjusting for BMI can qualitatively change our answer for methods that adjust for covariate error only, particularly for energy intake. The results from table \ref{supptablestrat} suggest that our results can change even more dramatically if we include BMI in the calibration model but exclude it from the outcome model. As an example, we see that this analysis approach results in a HR (95\% CI) for a 20\% increase in energy intake of 2.768 (2.279, 3.362), suggesting a much stronger association between intake and diabetes than seen previously. The results for protein and protein density intake also change substantially when BMI is included in the calibration model only.

Lastly, we were able to get similar results to \citet{tinker2011biomarker} by adopting a similar analysis approach and adding glycemic load, a covariate that was not in our calibration model, to our outcome model. In this case, the HR (95\% CI) for a 20\% increase in energy intake was 2.803 (2.314, 3.397) in the continuous model not adjusted for BMI. Finally, we note that discrepancies between results from our proposed approach and those of \citet{tinker2011biomarker} also stem from the fact that we have both corrected for outcome error and allowed for an imperfect specificity at baseline.

\newpage

\setcounter{table}{0}
\renewcommand{\thefigure}{S\arabic{figure}}

\begin{figure}[hbt!]
    \centering
    \includegraphics[width=1\textwidth]{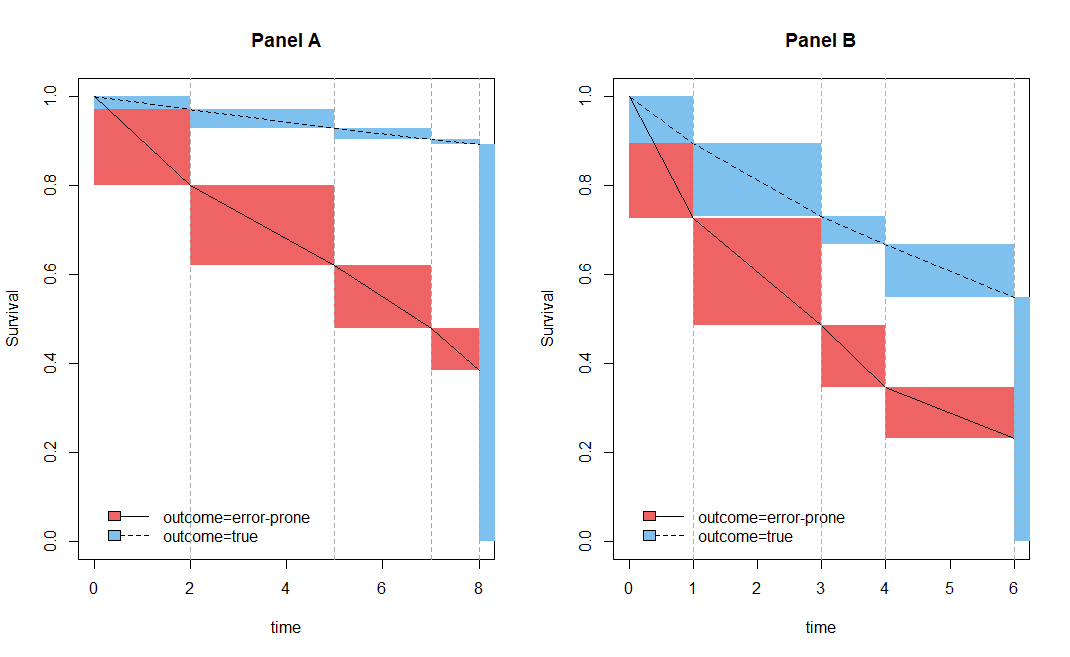}
    \caption{Estimated nonparametric maximum likelihood estimators (NPMLEs) of the survival distribution for the error-prone  outcomes compared to true outcomes for the simulation study, fit using the R package `interval.'\citep{fay2010exact} Panel A corresponds to censoring rate = 0.90 (baseline hazard = 0.012) with observation times $(2,5,7,8)$. Panel B corresponds to censoring rate = 0.55 (baseline hazard = 0.094) with observation times $(1,3,4,6)$. Vertical lines represent observation times. Simulated from data with $\beta_{X1}=\log(1.5)$, $\beta_{Z1}=\log(0.7)$, $\beta_{Z2}=\log(1.3)$, $e \sim \mathcal{N}(0,1.31)$, sensitivity = 0.90, and specificity = 0.80.}
    \label{fig:my_label}
\end{figure}

\renewcommand{\thetable}{S\arabic{table}}

\begin{table}
    \centering
      \begin{threeparttable}[t]
            \caption{The mean percent (\%) biases, average standard errors (ASE), empirical standard errors (ESE) and coverage probabilities (CP) are given for 1000 simulated data sets for the proposed method, the naive method, a method that corrects for covariate error only, and a method that corrects for outcome error only, with $\beta_{X1}=\log(1.5)$, $\beta_{Z1}=\log(0.7)$, and $\beta_{Z2}=\log(1.3)$; $e$ is normally distributed with mean zero; Sensitivity ($Se$)=0.80; Specificity ($Sp$)=0.90.}\label{table1simsappendix}
   
   \begin{tabular}{ccccccccccc}
\hline
\multicolumn{3}{l}{ } & \multicolumn{4}{c}{Proposed}  & \multicolumn{4}{c}{Naive} \\ \cmidrule(r){4-7} \cmidrule{8-11} $\hat{\delta}_{(1)}\tnote{1}$ & CR\tnote{2} & $\beta$ & \% Bias & ASE & ESE & CP & \% Bias & ASE & ESE & \multicolumn{1}{c}{CP} \\ 

\hline
0.60 & 0.90 & $\beta_{X1}$  & $\phantom{-}1.616$ & 0.200 & 0.204 & 0.950 & $-88.03$ & 0.046 & 0.046 & 0.000 \\ 
 &  & $\beta_{Z1}$  &  $-1.094$ & 0.143 & 0.142 & 0.945 & $-79.22$ & 0.057 & 0.058 & 0.002 \\ 
 &  & $\beta_{Z2}$  &  $-3.731$ & 0.143 & 0.143 & 0.945 & $-84.07$ & 0.057 & 0.054 & 0.021 \\ 
 & 0.55 & $\beta_{X1}$  &   $-1.231$ & 0.093 & 0.094 & 0.949 & $-68.11$ & 0.038 & 0.038 & 0.000 \\ 
 &  & $\beta_{Z1}$  & $-1.05$5 & 0.067 & 0.066 & 0.958 & $-43.46$ & 0.047 & 0.046 & 0.079 \\ 
 &  & $\beta_{Z2}$  & $-3.018$ & 0.066 & 0.065 & 0.957 & $-53.48$ & 0.046 & 0.045 & 0.133 \\ 
0.30 & 0.90 & $\beta_{X1}$  &  $\phantom{-}1.840$ & 0.283 & 0.286 & 0.954 & $-93.88$ & 0.033 & 0.033 & 0.000 \\ 
 &  & $\beta_{Z1}$  & $-1.233$ & 0.151 & 0.151 & 0.947 & $-82.46$ & 0.054 & 0.055 & 0.001 \\ 
 &  & $\beta_{Z2}$  & $-4.212$ & 0.151 & 0.150 & 0.945 & $-79.74$ & 0.054 & 0.052 & 0.025 \\ 
 & 0.55 & $\beta_{X1}$  & $-2.246$ & 0.131 & 0.133 & 0.940 & $-84.02$ & 0.027 & 0.027 & 0.000 \\ 
 &  & $\beta_{Z1}$  &  $-1.967$ & 0.071 & 0.069 & 0.951 & $-52.48$ & 0.045 & 0.044 & 0.008 \\ 
 &  & $\beta_{Z2}$  &  $-3.899$ & 0.070 & 0.068 & 0.956 & $-42.08$ & 0.045 & 0.044 & 0.306 \\ 
 \hline
\multicolumn{3}{l}{ } & \multicolumn{4}{c}{Correct Covariate Error}  & \multicolumn{4}{c}{Correct Outcome Error} \\ \cmidrule(r){4-7} \cmidrule{8-11} $\hat{\delta}_{(1)}$ & CR & $\beta$ & \% Bias & ASE & ESE & CP & \% Bias & ASE & ESE & \multicolumn{1}{c}{CP} \\ 

\hline
0.60 & 0.90 & $\beta_{X1}$  & $-80.15$ & 0.077 & 0.077 & 0.015 & $-38.84$ & 0.120 & 0.122 & 0.722 \\
 &  & $\beta_{Z1}$  &    $-80.62$ & 0.055 & 0.056 & 0.001 & $\phantom{-}6.019$ & 0.146 & 0.148 & 0.944 \\ 
 &  & $\beta_{Z2}$  &  $-82.19$ & 0.055 & 0.053 & 0.022 & $-13.31$ & 0.146 & 0.146 & 0.936 \\ 
 
 & 0.55 & $\beta_{X1}$  & $-47.05$ & 0.064 & 0.063 & 0.168 & $-40.51$ & 0.054 & 0.056 & 0.151 \\ 
 &  & $\beta_{Z1}$  & $-47.15$ & 0.046 & 0.045 & 0.042 & $\phantom{-}5.840$ & 0.067 & 0.068 & 0.942 \\
 &  & $\beta_{Z2}$  &   $-48.47$ & 0.046 & 0.044 & 0.192 & $-12.37$ & 0.066 & 0.066 & 0.919 \\  
0.30 & 0.90 & $\beta_{X1}$  & $-79.95$ & 0.109 & 0.108 & 0.150 & $-69.05$ & 0.085 & 0.086 & 0.109 \\ 
 &  & $\beta_{Z1}$  & $-80.62$ & 0.058 & 0.058 & 0.003 & $-10.77$ & 0.140 & 0.141 & 0.928 \\  
 &  & $\beta_{Z2}$  &  $-82.31$ & 0.058 & 0.056 & 0.030 & $\phantom{-}8.916$ & 0.140 & 0.140 & 0.947 \\ 
 & 0.55 & $\beta_{X1}$  & $-47.49$ & 0.091 & 0.089 & 0.419 & $-70.28$ & 0.038 & 0.040 & 0.000 \\ 
 &  & $\beta_{Z1}$  &  $-47.53$ & 0.049 & 0.047 & 0.059 & $-11.21$ & 0.064 & 0.064 & 0.892 \\ 
 &  & $\beta_{Z2}$  &  $-48.82$ & 0.048 & 0.046 & 0.231 & $\phantom{-}8.673$ & 0.064 & 0.064 & 0.938 \\ 
\hline
\multicolumn{3}{l}{} & \multicolumn{4}{c}{Truth}  \\ \cmidrule(r){4-7}   & CR & $\beta$ & \% Bias & ASE & ESE & CP \\
   \hline
  & 0.90 & $\beta_{X1}$ & 1.038 & 0.107 & 0.108 & 0.951 \\ 
  & & $\beta_{Z1}$ & 2.495 & 0.107 & 0.106 & 0.948  \\ 
  &   &$\beta_{Z2}$ & 2.444 & 0.107 & 0.108 & 0.948 \\ 
  & 0.55 &  $\beta_{X1}$  & 0.517 & 0.052 & 0.054 & 0.942 \\ 
  & &  $\beta_{Z1}$ & 1.471 & 0.052 & 0.053 & 0.951 \\ 
  &  &  $\beta_{Z2}$ & 1.773 & 0.052 & 0.052 & 0.948 \\ 
   \hline
\end{tabular}
 \begin{tablenotes}
     \item[1] $\hat{\delta}_{(1)}=$ estimate of attenuation coefficient \item[2] $CR=$ True censoring rate
   \end{tablenotes}
    \end{threeparttable}%
\end{table}

\begin{table}[ht]
\centering
\begin{threeparttable}[t]
\caption {The mean percent (\%) biases, average standard errors (ASE), empirical standard errors (ESE) and coverage probabilities (CP) are given for 1000 simulated data sets for the proposed method and naive method with $\beta_{X1}=\log(1.5)$, $\beta_{Z1}=\log(0.7)$, and $\beta_{Z2}=\log(1.3)$; $e$ is normally distributed with mean zero. The censoring rate is fixed at 0.90. Here, we vary sensitivity, specificity, and negative predictive value.} \label{supp2NPV} 
\begin{tabular}{lllcccccccc}
\hline
\multicolumn{3}{l}{ $Se\tnote{1} =0.80, Sp\tnote{2} =0.90$} & \multicolumn{4}{c}{Proposed}  & \multicolumn{4}{c}{Naive} \\ \cmidrule(r){4-7} \cmidrule{8-11} $\hat{\delta}_{(1)}\tnote{3}$ & $\eta$ \tnote{4} & $\beta$ & \% Bias & ASE & ESE & CP & \% Bias & ASE & ESE & \multicolumn{1}{c}{CP} \\ 

\hline
0.60 & 0.98 & $\beta_{X1}$  & $\phantom{-}3.401$ & 0.219 & 0.215 & 0.957 & $-88.00$ & 0.046 & 0.047 & 0.000 \\ 
 
 &  & $\beta_{Z1}$  &  $\phantom{-}3.644$ & 0.157 & 0.157 & 0.958 & $-79.01$ & 0.056 & 0.058 & 0.001 \\

 &  & $\beta_{Z2}$  & $\phantom{-}1.458$ & 0.156 & 0.160 & 0.946 & $-83.77$ & 0.056 & 0.058 & 0.028 \\ 
  
 & 0.90 & $\beta_{X1}$  & $\phantom{-}5.902$ & 0.270 & 0.275 & 0.951 & $-90.03$ & 0.043 & 0.044 & 0.000 \\ 
 &  & $\beta_{Z1}$  & $\phantom{-}4.952$ & 0.194 & 0.196 & 0.946 & $-82.51$ & 0.053 & 0.054 & 0.000 \\ 
  
 &  & $\beta_{Z2}$  & $-0.967$ & 0.191 & 0.199 & 0.947 & $-86.74$ & 0.053 & 0.054 & 0.015 \\ 
 
0.30 & 0.98 & $\beta_{X1}$  &  $\phantom{-}3.994$ & 0.311 & 0.300 & 0.960 & $-93.97$ & 0.033 & 0.033 & 0.000 \\ 
 &  & $\beta_{Z1}$  & $\phantom{-}3.631$ & 0.167 & 0.164 & 0.956 & $-82.33$ & 0.054 & 0.055 & 0.000 \\ 
 &  & $\beta_{Z2}$  & $\phantom{-}0.826$ & 0.165 & 0.169 & 0.945 & $-79.35$ & 0.054 & 0.056 & 0.032 \\ 
 & 0.90 & $\beta_{X1}$  & $\phantom{-}7.238$ & 0.384 & 0.383 & 0.963 & $-95.03$ & 0.031 & 0.031 & 0.000 \\ 
 &  & $\beta_{Z1}$  & $\phantom{-}5.193$ & 0.206 & 0.206 & 0.947 & $-85.29$ & 0.051 & 0.052 & 0.000 \\ 
 &  & $\beta_{Z2}$  & $-2.018$ & 0.203 & 0.210 & 0.950 & $-83.03$ & 0.051 & 0.052 & 0.018 \\ 
 \hline
\multicolumn{3}{l}{ $Se=0.90, Sp=0.80$} & \multicolumn{4}{c}{Proposed}  & \multicolumn{4}{c}{Naive} \\ \cmidrule(r){4-7} \cmidrule{8-11} $\hat{\delta}_{(1)}$ & $\eta$ & $\beta$ & \% Bias & ASE & ESE & CP & \% Bias & ASE & ESE & \multicolumn{1}{c}{CP} \\ 

\hline
0.60 & 0.98 & $\beta_{X1}$  &   $\phantom{-}2.100$ & 0.231 & 0.224 & 0.960 & $-93.58$ & 0.037 & 0.037 & 0.000 \\ 
 &  & $\beta_{Z1}$  & $\phantom{-}3.910$ & 0.166 & 0.163 & 0.957 & $-88.63$ & 0.046 & 0.046 & 0.000 \\ 
 &  & $\beta_{Z2}$  &  $\phantom{-}3.037$ & 0.164 & 0.167 & 0.949 & $-90.37$ & 0.045 & 0.046 & 0.000 \\ 
 & 0.90 & $\beta_{X1}$  & $\phantom{-}3.617$ & 0.283 & 0.285 & 0.956 & $-94.44$ & 0.036 & 0.036 & 0.000 \\ 
 &  & $\beta_{Z1}$  &  $\phantom{-}5.001$ & 0.203 & 0.207 & 0.939 & $-89.96$ & 0.045 & 0.045 & 0.000 \\ 
 &  & $\beta_{Z2}$  & $\phantom{-}1.572$ & 0.200 & 0.205 & 0.955 & $-91.46$ & 0.044 & 0.045 & 0.000 \\ 
0.30 & 0.98 & $\beta_{X1}$  & $\phantom{-}1.873$ & 0.327 & 0.316 & 0.965 & $-96.87$ & 0.026 & 0.027 & 0.000 \\ 
 &  & $\beta_{Z1}$  &  $\phantom{-}3.749$ & 0.175 & 0.171 & 0.954 & $-90.47$ & 0.044 & 0.044 & 0.000 \\ 
 &  & $\beta_{Z2}$  & $\phantom{-}2.754$ & 0.173 & 0.175 & 0.954 & $-87.93$ & 0.044 & 0.044 & 0.000 \\ 
 & 0.90 & $\beta_{X1}$  & $\phantom{-}3.600$ & 0.401 & 0.399 & 0.957 & $-97.33$ & 0.026 & 0.026 & 0.000 \\ 
 &  & $\beta_{Z1}$  & $\phantom{-}5.030$ & 0.215 & 0.216 & 0.942 & $-91.58$ & 0.043 & 0.044 & 0.000 \\ 
 &  & $\beta_{Z2}$  & $\phantom{-}1.134$ & 0.212 & 0.216 & 0.953 & $-89.30$ & 0.043 & 0.043 & 0.000 \\
 \hline
\multicolumn{3}{l}{ $Se=1, Sp=1, \eta=1$} & \multicolumn{4}{c}{Truth}  \\ \cmidrule(r){4-7}   &  & $\beta$ & \% Bias & ASE & ESE & CP \\
\hline 
 & & $\beta_1$ & $\phantom{-}1.962$ & 0.108 & 0.112 & 0.936 \\

 & & $\beta_2$ &  $\phantom{-}2.568$ & 0.107 & 0.109 & 0.944 \\ 
   & & $\beta_3$  &  $\phantom{-}1.115$ & 0.107 & 0.108 & 0.945 \\ 
  \hline
\end{tabular}
 \begin{tablenotes}
     \item[1] $Se=$ Sensitivity
     \item[2] $Sp=$ Specificity    
     \item[3] $\hat{\delta}_{(1)}=$ estimate of the attenuation coefficient
     \item[4] $\eta=$ Negative predictive value
   \end{tablenotes}
    \end{threeparttable}%
\end{table}

\begin{table}[ht]
\centering
\begin{threeparttable}[t]
\caption {The mean percent (\%) biases, average standard errors (ASE), empirical standard errors (ESE) and coverage probabilities (CP) are given for 1000 simulated data sets for the proposed method and naive method with $\beta_{X1}=\log(1.5)$, $\beta_{Z1}=\log(0.7)$, and $\beta_{Z2}=\log(1.3)$; $e$ is normally distributed with mean zero. The censoring rate is fixed at 0.90. Here, we vary sensitivity, specificity, and probability of missingness at each visit.} \label{supp3NPV} 
\begin{tabular}{lllcccccccc}
\hline
\multicolumn{3}{l}{ $Se\tnote{1} =0.80, Sp\tnote{2} =0.90$} & \multicolumn{4}{c}{Proposed}  & \multicolumn{4}{c}{Naive} \\ \cmidrule(r){4-7} \cmidrule{8-11} $\hat{\delta}_{(1)}\tnote{3}$ & $P_{Miss}$ \tnote{4} & $\beta$ & \% Bias & ASE & ESE & CP & \% Bias & ASE & ESE & \multicolumn{1}{c}{CP} \\ 

\hline
0.60 & 0.10 & $\beta_{X1}$  &   $-0.416$ & 0.206 & 0.205 & 0.952 & $-87.80$ & 0.048 & 0.049 & 0.000 \\
 &  & $\beta_{Z1}$  &  $-0.271$ & 0.148 & 0.152 & 0.943 & $-77.60$ & 0.059 & 0.062 & 0.004 \\ 
 
 &  & $\beta_{Z2}$  &    $-2.974$ & 0.148 & 0.154 & 0.945 & $-80.46$ & 0.059 & 0.061 & 0.052 \\ 
  
 & 0.40 & $\beta_{X1}$  & $-0.031$ & 0.243 & 0.244 & 0.955 & $-85.19$ & 0.056 & 0.057 & 0.000 \\ 
 
 &  & $\beta_{Z1}$  &    $\phantom{-}0.579$ & 0.173 & 0.177 & 0.940 & $-73.40$ & 0.068 & 0.071 & 0.034 \\ 
 
 &  & $\beta_{Z2}$  &   $-3.283$ & 0.173 & 0.180 & 0.942 & $-76.70$ & 0.068 & 0.071 & 0.168 \\ 
 
0.30 & 0.10 & $\beta_{X1}$  &  $-1.732$ & 0.292 & 0.292 & 0.952 & $-94.06$ & 0.034 & 0.034 & 0.000 \\ 
 
 &  & $\beta_{Z1}$  &  $-0.774$ & 0.156 & 0.160 & 0.954 & $-81.08$ & 0.056 & 0.059 & 0.001 \\ 
  
 &  & $\beta_{Z2}$  & $-2.701$ & 0.156 & 0.162 & 0.941 & $-75.82$ & 0.057 & 0.059 & 0.063 \\ 
   
 & 0.40 & $\beta_{X1}$  & $-1.297$ & 0.344 & 0.347 & 0.954 & $-92.82$ & 0.040 & 0.040 & 0.000 \\ 
  
 &  & $\beta_{Z1}$  &  $\phantom{-}0.036$ & 0.183 & 0.187 & 0.946 & $-77.63$ & 0.066 & 0.069 & 0.015 \\ 
  
 &  & $\beta_{Z2}$  &  $-3.146$ & 0.183 & 0.190 & 0.945 & $-71.04$ & 0.066 & 0.069 & 0.190 \\ 
 \hline
\multicolumn{3}{l}{ $Se=0.90, Sp=0.80$} & \multicolumn{4}{c}{Proposed}  & \multicolumn{4}{c}{Naive} \\ \cmidrule(r){4-7} \cmidrule{8-11} $\hat{\delta}_{(1)}$ & $P_{Miss}$ & $\beta$ & \% Bias & ASE & ESE & CP & \% Bias & ASE & ESE & \multicolumn{1}{c}{CP} \\ 

\hline
0.60 & 0.10 & $\beta_{X1}$  & $-1.920$ & 0.218 & 0.216 & 0.957 & $-93.38$ & 0.039 & 0.037 & 0.000 \\ 
 
 &  & $\beta_{Z1}$  &  $-0.451$ & 0.156 & 0.163 & 0.949 & $-87.87$ & 0.047 & 0.048 & 0.000 \\

 &  & $\beta_{Z2}$  &  $-2.801$ & 0.156 & 0.164 & 0.941 & $-89.04$ & 0.047 & 0.048 & 0.000 \\

 & 0.40 & $\beta_{X1}$  &    $-2.470$ & 0.264 & 0.268 & 0.958 & $-91.23$ & 0.044 & 0.044 & 0.000 \\ 
  
 &  & $\beta_{Z1}$  &  $-0.637$ & 0.189 & 0.200 & 0.944 & $-84.86$ & 0.054 & 0.056 & 0.000 \\

 &  & $\beta_{Z2}$  &    $-0.796$ & 0.189 & 0.200 & 0.946 & $-86.60$ & 0.054 & 0.055 & 0.012 \\

0.30 & 0.10 & $\beta_{X1}$  &    $-3.134$ & 0.308 & 0.309 & 0.953 & $-96.85$ & 0.028 & 0.026 & 0.000 \\ 

 &  & $\beta_{Z1}$  &   $-1.012$ & 0.165 & 0.171 & 0.953 & $-89.80$ & 0.045 & 0.046 & 0.000 \\ 
 
 &  & $\beta_{Z2}$  &   $-2.576$ & 0.165 & 0.174 & 0.944 & $-86.46$ & 0.045 & 0.046 & 0.000 \\ 
  
 & 0.40 & $\beta_{X1}$  &  $-4.353$ & 0.374 & 0.384 & 0.955 & $-95.77$ & 0.032 & 0.032 & 0.000 \\ 
  
 &  & $\beta_{Z1}$  & $-1.383$ & 0.200 & 0.211 & 0.944 & $-87.38$ & 0.052 & 0.054 & 0.000 \\ 
 
 &  & $\beta_{Z2}$  &   $-0.452$ & 0.200 & 0.213 & 0.945 & $-83.22$ & 0.052 & 0.053 & 0.015 \\   
 \hline
\multicolumn{3}{l}{ $Se=1, Sp=1, \eta=1$} & \multicolumn{4}{c}{Truth}  \\ \cmidrule(r){4-7}   & $P_{Miss}$ & $\beta$ & \% Bias & ASE & ESE & CP \\
\hline 
 & 0.10 & $\beta_1$ & $\phantom{-}0.848$ & 0.108 & 0.109 & 0.948 \\

 &  & $\beta_2$ &    $\phantom{-}1.458$ & 0.108 & 0.115 & 0.925 \\

   &  & $\beta_3$  &    $-1.882$ & 0.108 & 0.111 & 0.949 \\ 
    & 0.40 & $\beta_1$ & $\phantom{-}1.362$ & 0.114 & 0.117 & 0.943 \\ 
 
 &  & $\beta_2$ &    $\phantom{-}1.824$ & 0.114 & 0.123 & 0.929 \\

   &  & $\beta_3$  &   $-1.327$ & 0.114 & 0.117 & 0.956 \\  
  \hline
\end{tabular}
 \begin{tablenotes}
     \item[1] $Se=$ Sensitivity
     \item[2] $Sp=$ Specificity    
     \item[3] $\hat{\delta}_{(1)}=$ estimate of the attenuation coefficient
     \item[4] $P_{Miss}=$ Probability of missingness at each visit
   \end{tablenotes}
    \end{threeparttable}%
\end{table}

\begin{table}
    \centering

      \begin{threeparttable}[t]
            \caption{The mean percent (\%) biases, average standard errors (ASE), empirical standard errors (ESE) and coverage probabilities (CP) are given for 1000 simulated data sets for the proposed method, naive method, method that corrects for covariate error only, and method that corrects for outcome error only for a simulated dataset with similar features to the Women's Health Initiative (WHI) data. Here, Sensitivity ($Se$)=0.61, Specificity ($Sp$)=0.995, Negative Predictive Value ($\eta)=0.96$, $\beta_{X1}=\log(1.5)$, $\beta_{Z1}=\log(0.7)$, $\beta_{Z2}=\log(1.3)$, $e$ is normally distributed with mean zero, and the censoring rate for the error-prone indicator is fixed at 0.95.  }\label{WHIsims}
      \begin{tabular}{cccccccccc}
\hline
\multicolumn{2}{l}{ } & \multicolumn{4}{c}{Proposed}  & \multicolumn{4}{c}{Naive} \\ \cmidrule(r){3-6} \cmidrule{7-10} $\hat{\delta}_{(1)}\tnote{1}$ & $\beta$ & \% Bias & ASE & ESE & CP & \% Bias & ASE & ESE & \multicolumn{1}{c}{CP} \\ 

\hline
0.60 & $\beta_{X1}$ & $-0.294$ & 0.057 & 0.058 & 0.943 & $-79.21$ & 0.012 & 0.013 & 0.000 \\ 
 
 &  $\beta_{Z1}$  &      $\phantom{-}0.316$ & 0.041 & 0.042 & 0.939 & $-62.94$ & 0.015 & 0.015 & 0.000 \\ 
 
 &  $\beta_{Z2}$  &    $-0.578$ & 0.040 & 0.041 & 0.941 & $-68.75$ & 0.015 & 0.015 & 0.000 \\ 
 
0.30 & $\beta_{X1}$  &    $-0.186$ & 0.082 & 0.084 & 0.950 & $-89.53$ & 0.009 & 0.009 & 0.000 \\ 
 
 &  $\beta_{Z1}$  &     $\phantom{-}0.366$ & 0.044 & 0.046 & 0.940 & $-68.63$ & 0.014 & 0.015 & 0.000 \\ 
  
 &  $\beta_{Z2}$  &   $-0.786$ & 0.044 & 0.044 & 0.941 & $-61.04$ &    0.014 & 0.015 & 0.000 \\ 
 \hline
\multicolumn{2}{l}{ } & \multicolumn{4}{c}{Correct Covariate Error}  & \multicolumn{4}{c}{Correct Outcome Error} \\ \cmidrule(r){3-6} \cmidrule{7-10} $\hat{\delta}_{(1)}$ & $\beta$ & \% Bias & ASE & ESE & CP & \% Bias & ASE & ESE & \multicolumn{1}{c}{CP} \\ 
\hline
0.60 & $\beta_{X1}$ & $-65.51$ & 0.022 & 0.022 & 0.000 & $-39.92$ & 0.032 & 0.032 & 0.000 \\ 
  
 &  $\beta_{Z1}$  &     $-65.26$ & 0.015 & 0.016 & 0.000 & $\phantom{-}7.018$ & 0.039 & 0.040 & 0.902 \\ 
 
 &  $\beta_{Z2}$  &      $-65.55$ & 0.015 & 0.016 & 0.000 & $-9.828$ & 0.039 & 0.039 & 0.896 \\ 
0.30 & $\beta_{X1}$  &     $-65.48$ & 0.031 & 0.032 & 0.000 & $-69.73$ & 0.023 & 0.023 & 0.000 \\ 
  
 &  $\beta_{Z1}$  &     $-65.25$ & 0.017 & 0.017 & 0.000 & $-9.444$ & 0.038 & 0.038 & 0.853 \\ 
  
 &  $\beta_{Z2}$  &     $-65.61$ & 0.017 & 0.017 & 0.000 & $\phantom{-}12.408$ & 0.038 & 0.037 & 0.867 \\ 
\hline
\multicolumn{2}{l}{} & \multicolumn{4}{c}{Truth}  \\ \cmidrule(r){3-6}  & $\beta$ & \% Bias & ASE & ESE & CP \\
   \hline
  & $\beta_{X1}$ &    $-0.208$ & 0.022 & 0.022 & 0.961 \\

    & $\beta_{Z1}$ &   $\phantom{-}0.026$ & 0.022 & 0.023 & 0.952 \\ 
 
  &  $\beta_{Z2}$ &       $\phantom{-}0.200$ & 0.022 & 0.022 & 0.947 \\ 
   \hline
\end{tabular}
 \begin{tablenotes}
     \item[1] $\hat{\delta}_{(1)}=$ estimate of attenuation coefficient 
   \end{tablenotes}
    \end{threeparttable}%
\end{table}
% latex table generated in R 3.5.2 by xtable 1.8-3 package
% Thu Sep 12 11:45:50 2019
\begin{table}[ht]
 \begin{threeparttable}[t]

\caption {Hazard Ratio (HR) and 95\% confidence interval (CI) estimates of incident diabetes for a 20\% increase in consumption of energy (kcal/d), protein (g/d), and protein density (\% energy from protein/d) based on the naive method ignoring error in the outcome and covariate, the method corrected for error in the covariate only, and the proposed method. Here, sensitivity = 0.61, specificity = 0.995, and negative predictive value = 1.} \label{suppWHI} 

\centering
\begin{tabular}{llll}
  \hline
  \multicolumn{2}{c}{} & \multicolumn{2}{c}{HR (95\% CI)} \\  \multicolumn{1}{l}{Model\tnote{1}} & \multicolumn{1}{l}{Method} & \multicolumn{1}{c}{Adjusted for BMI\tnote{2}}  & \multicolumn{1}{c}{Not Adjusted for BMI} \\ 
  \hline
 Energy (kcal/d) & 

  Naive & 1.002 (0.986, 1.018) & 1.024 (1.008, 1.040) \\ 
  & Regression Calibration & 1.041 (0.758, 1.429) & 1.421 (1.043, 1.938) \\ 
  & Proposed & 0.973 (0.714, 1.327) & 1.314 (0.992, 1.740) \\   \hline

   Protein (g/d) & Naive & 1.024 (1.010, 1.039) & 1.051 (1.035, 1.066) \\ 
  & Regression Calibration & 1.121 (1.036, 1.213) & 1.231 (1.130, 1.342) \\ 
  & Proposed & 1.107 (1.025, 1.195) & 1.229 (1.128, 1.339) \\   \hline

  Protein Density & Naive & 1.100 (1.064, 1.137) & 1.128 (1.091, 1.167) \\ 
  & Regression Calibration & 1.243 (1.125, 1.374) & 1.325 (1.181, 1.486) \\ 
  & Proposed & 1.209 (1.100, 1.329) & 1.327 (1.183, 1.490) \\ 
   \hline
\end{tabular}
\begin{tablenotes}
     \item[1] Each model is adjusted for potential confounders and is stratified on age (10-year categories) and Dietary Modification trial (DM) or Observational Study (OS) cohort membership.
     \item[2] BMI = Body Mass Index $(kg/m^2)$
   \end{tablenotes}
       \end{threeparttable}%
\end{table}

% latex table generated in R 3.5.2 by xtable 1.8-3 package
% Fri Sep 20 09:46:45 2019
\begin{table}[ht]
\centering
\caption {Sensitivity analysis varying sensitivity and specificity of diabetes self-reports across WHI DM-C and WHI OS participants. We consider separate models for dietary energy, protein, and protein density. Each model is adjusted for potential confounders, including BMI, and is stratified on age (10-year categories) and DM or OS cohort membership. We show HR estimates of incident diabetes for a 20\% increase in consumption of energy (kcal/d), protein (g/d), and protein density (\% energy from protein/d).} \label{supp1} 
\scalebox{0.9}{\begin{tabular}{lllllll}
  \hline
\multicolumn{2}{c}{Sensitivity} & \multicolumn{2}{c}{Specificity} & \multicolumn{3}{c}{HR (95\% CI)} \\
OS &  DM & OS & DM & Energy (kcal/d)  & Protein (g/d)  & Protein Density \\ 
  \hline
0.5800 & 0.7418 & 0.9945 & 0.9972 & 0.970 (0.713,1.319) & 1.113 (1.030,1.202) & 1.193 (1.088,1.307) \\ 
  0.5300 & 0.9614 & 0.9945 & 0.9972 & 0.954 (0.699,1.302) & 1.114 (1.031,1.203) & 1.183 (1.081,1.295) \\ 
  0.6168 & 0.5800 & 0.9945 & 0.9972 & 0.938 (0.690,1.276) & 1.108 (1.027,1.195) & 1.199 (1.093,1.314) \\ 
  0.6282 & 0.5300 & 0.9945 & 0.9972 & 0.959 (0.700,1.313) & 1.106 (1.025,1.194) & 1.183 (1.081,1.293) \\ 
  0.5800 & 0.7418 & 0.9951 & 0.9945 & 0.974 (0.715,1.326) & 1.108 (1.026,1.196) & 1.206 (1.099,1.324) \\ 
  0.5300 & 0.9614 & 0.9951 & 0.9945 & 0.971 (0.710,1.327) & 1.110 (1.028,1.199) & 1.193 (1.088,1.308) \\ 
  0.6168 & 0.5800 & 0.9951 & 0.9945 & 0.972 (0.709,1.333) & 1.105 (1.025,1.191) & 1.173 (1.074,1.282) \\ 
  0.6282 & 0.5300 & 0.9951 & 0.9945 & 0.960 (0.705,1.306) & 1.108 (1.027,1.195) & 1.189 (1.087,1.302) \\ 
   \hline
\end{tabular}}
\end{table}

\begin{table}
 \begin{threeparttable}[t]
\caption {Sensitivity Analysis for different stratification strategies using a modeling approach similar to that of Tinker et al. \cite{tinker2011biomarker} We examine hazard ratio (HR) and 95\% confidence interval (CI) estimates of incident diabetes for a 20\% increase in consumption of energy (kcal/d), protein (g/d), and protein density (\% energy from protein/d) based on discrete proportional hazards analyses and continuous Cox proportional hazards models that correct for error in the covariate ($X$) only. } \label{stratsens} 

\centering
\begin{tabular}{llll}
  \hline
   \multicolumn{2}{c}{} & \multicolumn{2}{c}{HR (95\% CI)} \\  \multicolumn{1}{l}{Model\tnote{1}} & \multicolumn{1}{l}{Method} & \multicolumn{1}{c}{Adjusted for BMI\tnote{2}}  & \multicolumn{1}{c}{Not Adjusted for BMI} \\
  \hline
 Energy (kcal/d) & 
  Discrete, 6 Strata\tnote{3} & 1.041 (0.758, 1.429) & 1.421 (1.043, 1.938) \\ 
 & Continuous, 6 Strata & 0.953 (0.686, 1.323) & 1.333 (0.993, 1.790) \\ 
 & Continuous, 12 Strata\tnote{4} & 0.953 (0.686, 1.324)
 & 1.334 (0.994, 1.791)
 \\ 
 & Continuous, 36 Strata\tnote{5} & 0.952 (0.685, 1.322)
 & 1.328 (0.990, 1.780)
 \\  \hline

   Protein (g/d) & 
  Discrete, 6 Strata & 1.121 (1.036, 1.213)

 & 1.231 (1.130, 1.342)

 \\ 
  & Continuous, 6 Strata & 1.104 (1.020, 1.194)

 & 1.217 (1.117, 1.325)

 \\ 
  & Continuous, 12 Strata & 1.103 (1.020, 1.193)

 & 1.216 (1.117, 1.324)

 \\ 
& Continuous, 36 Strata & 1.104 (1.021, 1.194)

 & 1.215 (1.116, 1.323)

 \\  \hline

  Protein Density & 
 Discrete, 6 Strata & 1.243 (1.125, 1.374) &	1.325 (1.181, 1.486)
 \\ 
 & Continuous, 6 Strata & 1.241 (1.121, 1.374)
 & 1.325 (1.179, 1.489)
 \\ 
  & Continuous, 12 Strata & 1.241 (1.122, 1.374)
 & 1.324 (1.179, 1.487)
 \\ 
  & Continuous, 36 Strata & 1.243 (1.123, 1.377)
 & 1.324 (1.179, 1.487)
 \\  \hline

   \hline
\end{tabular}
 \begin{tablenotes}
     \item[1] Each model is adjusted for potential confounders
     \item[2] BMI = Body Mass Index $(kg/m^2)$
     \item[3] 6 strata: age (10-year categories) and Dietary Modification trial (DM) or Observational Study (OS) cohort membership
     \item[4] 12 strata: age (5-year categories) and Dietary Modification trial (DM) or Observational Study (OS) cohort membership
     \item[5] 36 strata: age (5-year categories), Dietary Modification trial (DM) or Observational Study (OS) cohort membership, and hormone therapy trial arm. 
   \end{tablenotes}
    \end{threeparttable}%
\end{table}

\begin{table}
\begin{adjustbox}{max width=\textwidth}
 \begin{threeparttable}[ht]
\caption {For model used by Tinker et al., \cite{tinker2011biomarker}, we examine the sensitivity of results to choices of how BMI is treated in analyses. We present hazard ratio (HR) and 95\% confidence interval (CI) estimates of incident diabetes for a 20\% increase in consumption of energy (kcal/d), protein (g/d), and protein density (\% energy from protein/d) based on the discrete proportional hazards model with a post-hoc correction for covariate error, the continuous Cox model with a post-hoc correction for covariate error, and the continuous Cox model with the non-post-hoc traditional regression calibration correction for covariate error.} \label{supptablestrat} 

\centering
\begin{tabular}{lllll}
  \hline
   \multicolumn{2}{c}{} & \multicolumn{3}{c}{HR (95\% CI)} \\  \multicolumn{1}{l}{Model\tnote{1}} & \multicolumn{1}{l}{Method} & \multicolumn{1}{c}{BMI in Both\tnote{2}}  & \multicolumn{1}{c}{BMI in Neither\tnote{3}} & \multicolumn{1}{c}{Calibration Only\tnote{4}} \\
  \hline
 Energy (kcal/d) &  
 
  Discrete Post-Hoc	& 1.041 (0.758, 1.429) &	1.421 (1.043, 1.938) &	NA \\
  & Continuous Post-Hoc &	0.953 (0.686, 1.323)	& 1.333 (0.993, 1.790) &	NA \\
& Continuous Non-Post-Hoc 	& 0.956 (0.650, 1.407) & 1.290 (0.967, 1.720) &	2.768 (2.279, 3.362) \\ \hline

   Protein (g/d) & 
Discrete, Post-Hoc & 1.121 (1.036, 1.213) &	1.231 (1.130, 1.342) & NA
 \\
   & Continuous Post-Hoc  &	1.104 (1.020, 1.194)	& 1.217 (1.117, 1.325) &	NA \\
& Continuous Non-Post-Hoc 	& 1.099 (1.009, 1.196)
	&  1.208 (1.095, 1.333)
&	1.790 (1.430, 2.242) \\ \hline

  Protein Density & 
  Discrete Post-Hoc	& 1.243 (1.125, 1.374) &	1.325 (1.181, 1.486) &	NA \\
   & Continuous Post-Hoc  &	1.241 (1.121, 1.374) &	1.325 (1.179, 1.489) &	NA \\
& Continuous Non-Post-Hoc & 1.226 (1.111, 1.352)
 & 1.303 (1.161, 1.463)
	&  1.049 (0.689, 1.597)
 \\ \hline

   \hline
\end{tabular}
 \begin{tablenotes}
     \item[1] Each model is adjusted for potential confounders and stratified on age (10-year categories) and Dietary Modification trial (DM) or Observational Study (OS) cohort membership.
     \item [2] BMI included in both calibration and outcome model (BMI = Body Mass Index $(kg/m^2)$)
    \item[3] BMI included in neither the calibration nor the outcome model 
     \item[4] BMI included in the calibration model but not the outcome model

   \end{tablenotes}
    \end{threeparttable}
    \end{adjustbox}

\end{table}

\label{lastpage}

\end{document}